\documentclass{aa}  

\usepackage{graphicx}
\usepackage{txfonts}
\usepackage{amssymb}
\usepackage{mathtools}
\usepackage{lscape}

\usepackage{bm}

\usepackage{float}
\usepackage{graphicx}
\usepackage{txfonts}
\usepackage{gensymb}
\usepackage{adjustbox}
\usepackage{multirow}
\usepackage{epstopdf}
\usepackage{tabulary}
\usepackage{tabularx}
\usepackage{longtable}
\usepackage{hhline}
\usepackage{booktabs}
\usepackage{array}
\usepackage{subcaption}
\usepackage{titlesec}
\usepackage[normalem]{ulem}

\begin{document}

   \title{Regolith behavior under asteroid-level gravity conditions: Low-velocity impacts into mm- and cm-sized grain targets}

   \author{J. Brisset\inst{1}
          \and
          C. Cox\inst{2}
          \and
          S. Anderson\inst{2}
          \and
          J. Hatchitt\inst{1}
          \and 
          A. Madison\inst{1}
          \and
          M. Mendonca\inst{1}
          \and
          A. Partida\inst{1}
          \and
          D. Remie\inst{1}
          }

   \institute{Florida Space Institute, University of Central Florida, 12354 Research Parkway, Partnership 1 Building, Suite 214, Orlando, FL 32826-0650\\
              \email{julie.brisset@ucf.edu}
         \and
             Physics Department, University of Central Florida, 4111 Libra Drive, Orlando, FL 32816
             }

%   \date{XXX}

% \abstract{}{}{}{}{} 
% 5 {} token are mandatory
 
  \abstract
  % context heading (optional)
  % {} leave it empty if necessary  
   {In situ observations of small asteroids, such as Itokawa, Ryugu, and Bennu, show that surfaces covered by boulders and coarse terrain are frequent on such bodies. Regolith grain sizes have distributions on approximately mm and cm scales, and the behavior of such large grains in the very low-gravity environments of small body surfaces dictates their morphology and evolution.}
  % aims heading (mandatory)
   {In order to support the understanding of natural processes (e.g., the recapturing of impact ejecta) or spacecraft-induced interactions (e.g., the fate of a small lander), we aim to experimentally investigate the response of coarse-grained target surfaces to very-low-speed impacts (below 2~m/s).}
  % methods heading (mandatory)
   {We present the outcome of 86 low-speed impacts of a cm-sized spherical projectile into a bed of simulated regolith, composed of irregular mm- and cm-sized grains. These impacts were performed under vacuum and microgravity conditions. Our results include measurements for the projectile coefficient of restitution and penetration depth, as well as ejecta production, speed, and mass estimation. As part of our data analysis, we compared our data set with impacts performed in similar conditions with fine grain regolith targets to determine the dependence of our measurements on the target grain size.}
  % results heading (mandatory)
   {We find that impact outcomes include the frequent occurrence of projectile bouncing and tangential rolling on the target surface upon impact. Ejecta is produced for impact speeds higher than about 12~cm/s, and ejecta speeds scale with the projectile to target the grain size ratio and the impact speed. Ejected mass estimations indicate that ejecta is increasingly difficult to produce for increasing grain sizes. Coefficients of restitution of rebounding projectiles do not display a dependency on the target grain size, unlike their maximum penetration depth, which can be scaled with the projectile to target grain size ratio. Finally, we compare our experimental measurements to spacecraft data and numerical work on Hayabusa 2's MASCOT landing on the surface of the asteroid Ryugu.}
  % conclusions heading (optional), leave it empty if necessary 
   {}


   \maketitle
%
%-------------------------------------------------------------------

\section{Introduction}
\label{intro}

Data returned from spacecraft missions to asteroids of the Solar System increasingly show the ubiquity of rough surfaces covered in coarse grains and boulders. While some asteroids, such as Eros and Itokawa, display a variety of surface features with smooth ponds alternating with bouldered terrain \citep{fujiwara2006rubble,saito2006detailed}, more recent and detailed data from JAXA's Hayabusa2 at (162173)Ryugu and NASA's OSIRIS-REx at (101955)Bennu show surfaces that are entirely rubble-covered and coarse \citep{lauretta2019unexpected,watanabe2019hayabusa2}. Coexisting with large boulders ($\sim$10~m), regolith grain sizes on these bodies are estimated to range from mm to dm \citep{sugita2019geomorphology}. As such surfaces appear to be common among small asteroids, understanding how such coarse-grained surfaces behave in asteroid-like environments can shed light on the observed topography, activity \citep{lauretta2019episodes}, and impact response of these bodies.

The discovery of rubbled terrain on asteroid (25143)Itokawa in 2005 triggered experimental and numerical work on impacts into coarse and boulder-covered regolith terrain in an effort to understand the terrain's influence on cratering and ejecta processes. In particular, the absence of small ($<$ 100~m) craters on such coarse terrain \citep{chapman2002impact,hirata2009survey} is indicative of either modified impact dynamics compared to fine grain surfaces, or post-impact processes leading to their suppression \citep[e.g., impact-induced seismic activity, ][]{richardson2004impact}. For example, the "armoring effect," in which a larger boulder lying on a regolith surface is the primary contact of an incoming impactor and absorbs part of the impact energy, has been studied in several experimental investigations \citep{durda2011experimental,guttler2012cratering,tatsumi2018cratering,barnouin2019impacts}. In addition, experiments on coarse-grained targets have been conducted, showing that the ratio of target grain size to projectile size plays a role in cratering processes \citep{cintala1999ejection,guttler2012cratering,barnouin2019impacts,tatsumi2018cratering}. These experiments commonly used mm- to cm-sized grains both as target materials and as projectiles. Impact speeds ranged from 80~m/s to a few km/s.

In the work presented here, we investigate impacts into coarse grain targets at speed ranges from 0.1 to 2~m/s, several orders of magnitude smaller than these previous experiments. These speed ranges apply to both natural and spacecraft-induced processes. Indeed, recaptured secondary ejecta, which has speeds below the body's escape speed, is expected to impact the surface of a small asteroid at speeds around or below 25~cm/s (26~cm/s for Ryugu, 12~cm/s for Itokawa) and a few m/s for somewhat larger asteroids (e.g., 7.3~m/s for (433)Eros). Such low impact speeds also apply to re-captured lifted particles, as seen in the activity at the surface of asteroid Bennu \citep{lauretta2019episodes}. In addition, recent and current missions to asteroids have interacted or plan to interact with their coarse-grain surfaces at speeds around or below 20~cm/s in order to land hardware or collect surface samples. For example, Hayabusa2's Mobile Asteroid Surface Scout (MASCOT) lander touched down on Ryugu's surface at a speed of 17~cm/s in October 2018 \citep{scholten2019descent} and the Touch-And-Go (TAG) operation of the OSIRIS-REx mission, planned for October 2020, will make contact with Bennu's surface at around 10~cm/s \citep{lauretta2017osiris}. Understanding impact processes and surface responses at these low speeds can provide much needed context for surface interaction activities on these small bodies, and thus increase the mission's science return. While numerical work has been performed in support of these mission operations \citep{maurel2018numerical,thuillet2018numerical,thuillet2019numerical}, no microgravity experimental measurements of impacts into coarse material at such low impact speeds are yet available.

In \cite{brisset2018regolith}, we presented experimental results from low-speed impacts into simulated fine-grained regolith ($<$ 0.25~mm). In order to achieve speeds below 1 m/s and observe accurate target responses, these experiments have to be performed under microgravity conditions. For example, to generate an impact at 5 cm/s, an aggregate would have to be dropped from a height of 130 $\mu$m, and its free-fall time would be 13 $\mu$s. Such experiment parameters make it impossible to properly record the impact. In addition, the acceleration of gravity on Earth dominates target grain behavior at such low impact energies, thus masking the ejecta behavior. In \cite{brisset2018regolith}, we showed that the target response in particular is sensitive to the quality of the microgravity environment in which the experiments are performed, with a target behavior transition at about 10$^{-3}g$, $g$ being the Earth acceleration of gravity. In the present work, we study impacts into coarse grains (mm- to cm-size ranges) using a laboratory drop tower, in which a 10$^{-4}g$ microgravity environment is reproduced. The hardware setup, projectiles, and impact speeds were kept similar to experiments performed in \cite{brisset2018regolith}, allowing for a comparison of the two data sets and a measurement of the influence of target grain size on the impact outcome. The study of larger grain sizes allows us to investigate the competition between gravitational and cohesive forces between regolith grains at the surface of small
asteroids. Coarse grains are also relevant to the the study of surface processes on small bodies such as Ryugu and Bennu, as recent spacecraft data indicates a depletion in fine grains on these asteroids \citep[Hayabusa2, ][ and OSIRIS-REx, \citealt{dellagiustina2019properties}]{sugita2019geomorphology}. In Section~\ref{s:setup}, we describe our experimental setup as well as the various impact parameters explored. In Section~\ref{s:analysis}, we present our data analysis methods used to obtain the data results presented in Section~\ref{s:results}. We present some discussion points on these results in Section~\ref{s:discussion}. Finally, we summarize our findings in Section~\ref{s:conclusion}.

\begin{figure}[t]
  \begin{center}
  \includegraphics[width = 0.45\textwidth]{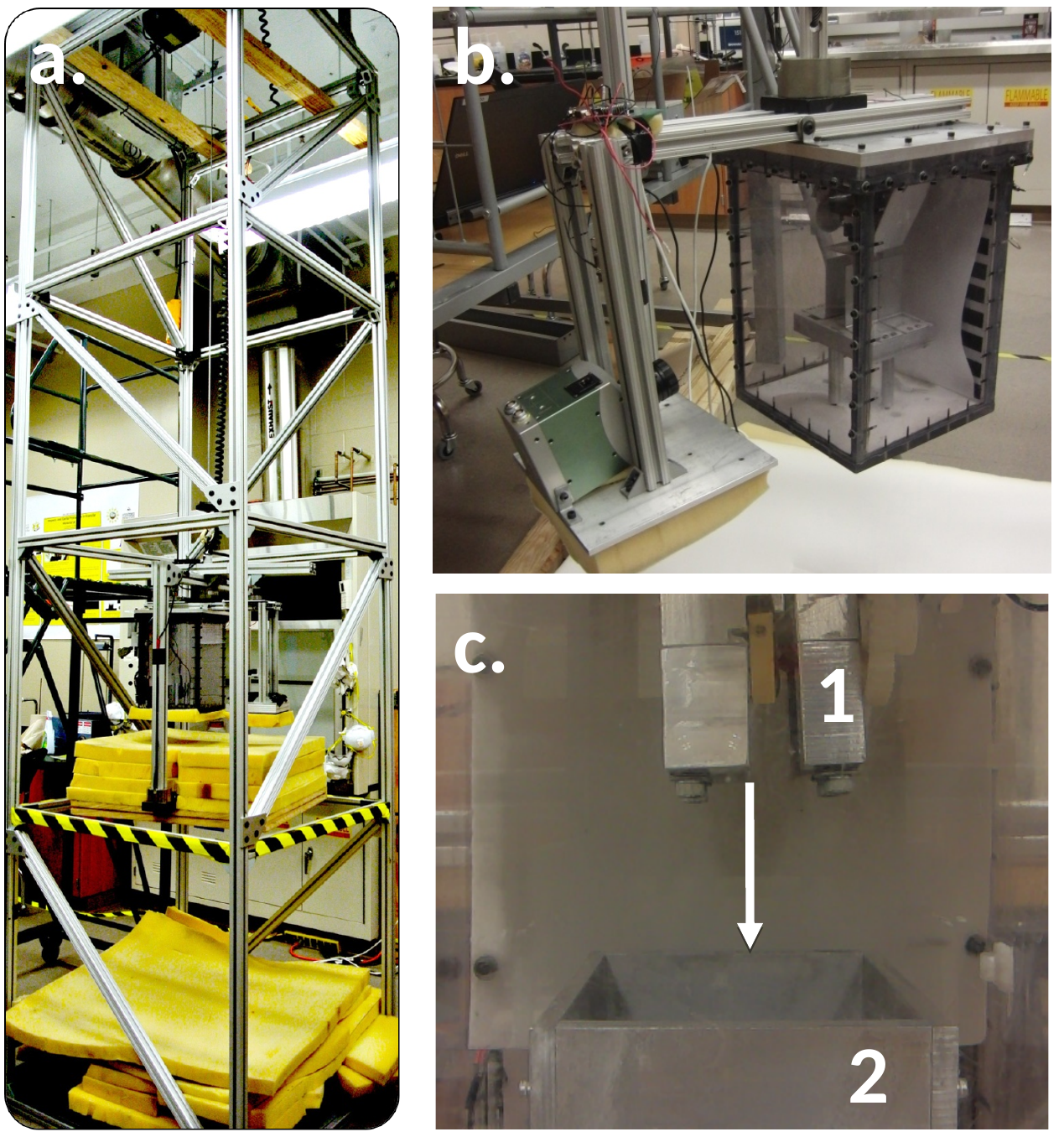}
 \caption{PRIME-D drop-tower experimental setup: (a) laboratory drop tower providing for 0.75~s of free-fall time; (b) drop-tower experiment box with camera attached; (c) hardware components inside the PRIME-D experiment box. The arrow indicates the travel path of the projectile after departure from the launch mechanism (1) toward the sample tray (2) located at the bottom of the picture.}
 \label{f:HW}
 \end{center}
\end{figure}

%--------------------------------------------------------------------
\section{Experiment setup}
\label{s:setup}

In this section, we describe the experiment hardware and target and projectile material, as well as the experiment plan implemented for data collection.

\subsection{Experiment hardware}
\label{s:HW}

\begin{figure*}[ht]
    \centering
    \begin{minipage}{0.49\textwidth}
        \centering
        \includegraphics[width=0.9\textwidth]{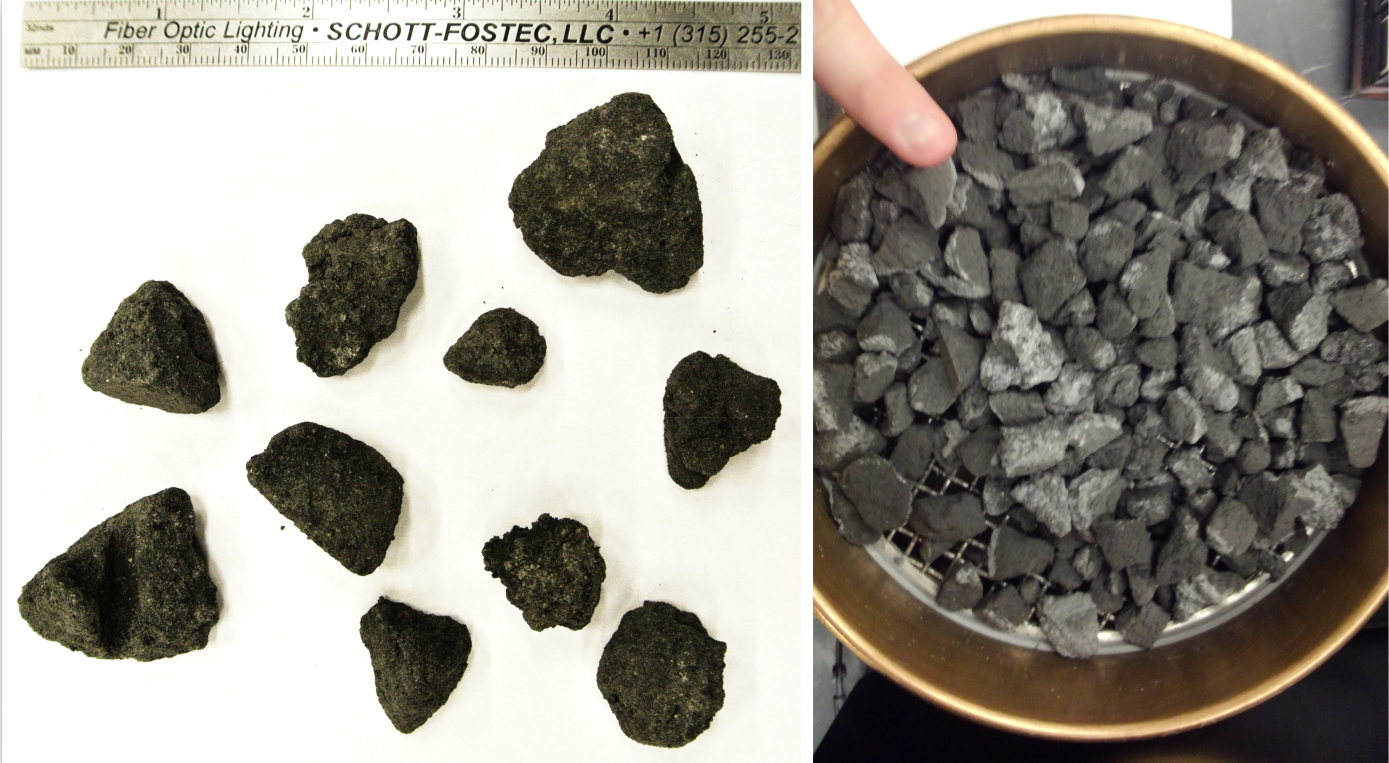} 
    \end{minipage}\hfill
    \begin{minipage}{0.49\textwidth}
        \centering
        \includegraphics[width=0.9\textwidth]{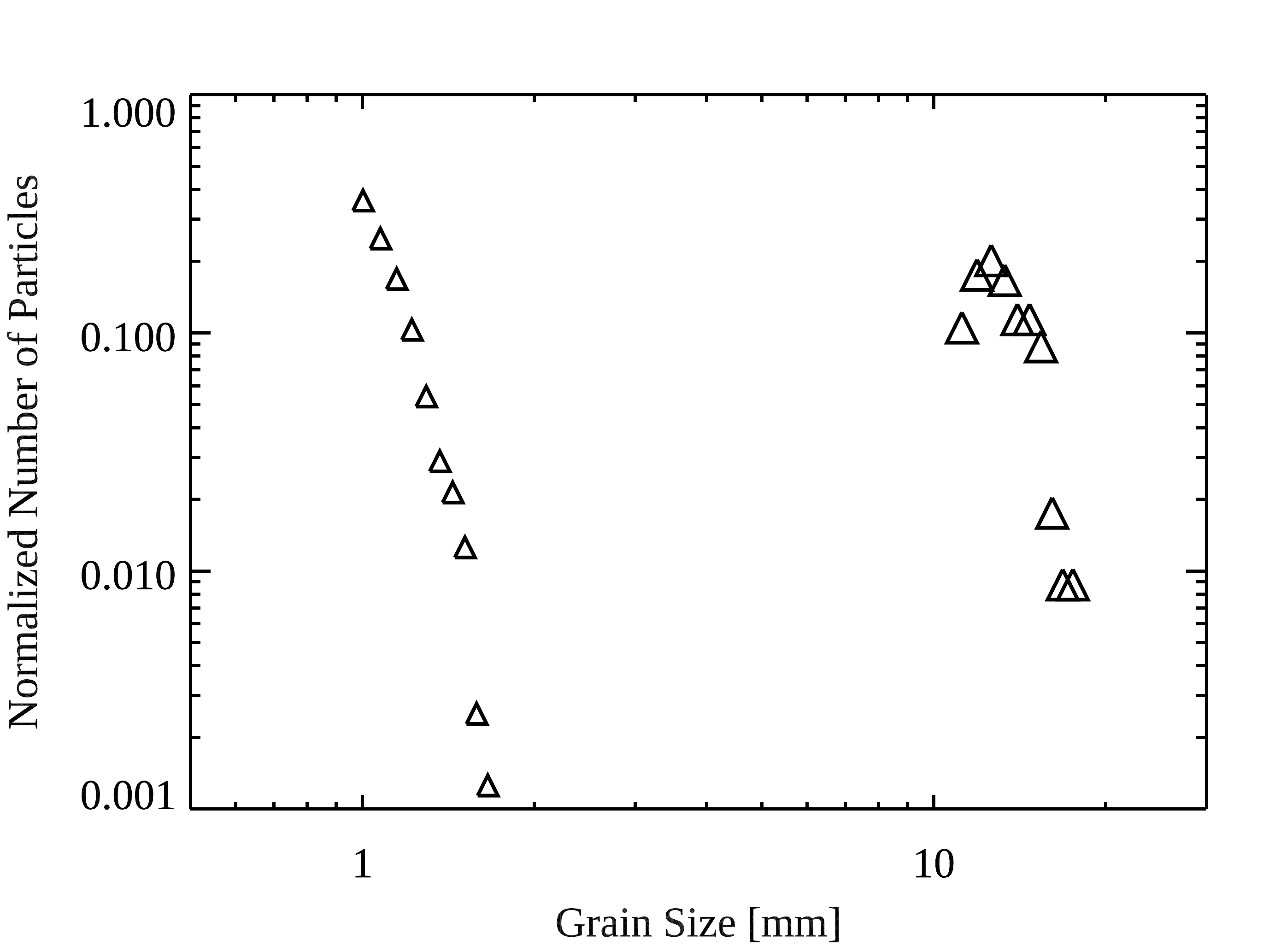} 
    \end{minipage}
    \caption{Sample material grains: (left) cm-sized grains produced from UCF-1 CI Orgueil simulant. Millimeter-sized grains are very similar in shape, surface roughness, and overall aspect of the grains with the only difference being the smaller grain size; (right) grain size distributions of the two populations prepared. The mm grain population is shown with the smaller symbols and peaks at 1~mm, while the cm population is shown with the larger symbols and peaks at about 13~mm. Both distributions were normalized with the total number of grains considered for the measurement (over 1,000 for mm-sized grains and over 100 for cm-sized grains).}
    \label{f:grains}
\end{figure*}

The impact data presented here were collected with the Physics of Regolith Impacts in Microgravity Experiment - Drop tower (PRIME-D). The PRIME-D hardware design relies on the same concept as the PRIME payload flown on parabolic aircraft \citep{brisset2018regolith}; inside a vacuum box, a projectile is launched into a regolith sample tray of dimensions 13$\times$13$\times$3 cm$^3,$ while the whole assembly is under free-fall. This tray can be removed and the whole bottom half of the vacuum box (for a target volume of 30$\times$30$\times$40 cm$^3$) can be filled with target material, which was performed in order to test the influence of the target tray size on the impact outcome. The projectile launch mechanism is composed of a shim holding the projectile, which is positioned on a compressed spring. The shim and projectile are held in place by a solenoid mechanism. Upon release, the compressed spring pushes the shim, and thus the projectile, normally toward the target. The final launch speed is therefore dependent on the projectile and shim masses as well as the spring constant. The sample tray is located under the projectile launching mechanism for normal impacts (see Figure~\ref{f:HW},c).

The PRIME-D vacuum box (Figure~\ref{f:HW},b) is composed of acrylic side walls so that a camera attached to the exterior of the box can record the impact. The target sample tray is fixed to the bottom plate of the box, while the launch mechanism is attached to the top plate. The camera recorded the impacts with 1020p resolution at 500 fps. An LED panel on the opposite side of the target tray provided back-lighting for the camera recording, thus easing the projectile and ejecta tracking process. 

The vacuum levels routinely achieved in the PRIME-D vacuum box are around a few tens of mTorr. While this is much higher than the ambient pressure to be expected at the surface or around asteroids, it achieves two main goals: eliminating the viscous fluid effects of air on the particle trajectory and reducing the cohesion induced by the presence of water vapor in the ambient gas, which impacts the cohesion between target grains. Previous experiments on dust grains and regolith in vacuum \citep{brisset2016ICE,brisset2018regolith} have shown that vacuum levels $\leq$ 100 mTorr are sufficient to satisfy these goals and thus achieve satisfactory conditions to study the impact behavior of simulated regolith targets.

For the PRIME-D impacts presented here, the microgravity environment was obtained via the use of a laboratory drop tower (Figure~\ref{f:HW},a). With a height of 3.4 m, this drop tower allows for 0.75 s of microgravity time. The experiment box and camera assembly is equipped with a magnet that attaches to a powered counterpart at the top of the drop tower. When power is cut to the tower top magnet, the box and camera assembly detaches and drops onto several layers of damping foam. A lead drag shield dropped under the experiment box a fraction of a second earlier allows for a microgravity quality of about 10$^{-4}g$. The impact speeds achieved with this PRIME-D setup range from 7~cm/s to about 2 m/s (see Table~\ref{t:runs}).

\subsection{Target material}

In the present work, we used UCF-1 CI Orgueil asteroid regolith simulant to prepare target material grains \citep{covey2016simulating}. This simulant comes as powder of particles of sizes $<$ 100~$\mu$m. Mixed with water, dried out in large ($>$10~cm) chunks, and smashed with a hammer, this powder can be prepared as grains of any size (Figure~\ref{f:grains}). In order to extend the parameter space investigated in \cite{brisset2018regolith}, we focused on larger grains in the mm and cm ranges. The rationale for this choice is two-fold: first, one of the goals of the present experiments is to study the competition between gravitational and cohesive forces between simulated regolith grains at the surface of small asteroids. As shown in \cite{brisset2018regolith} (their Figure~13), the transition between a gravity- and cohesion-dominated regime for small asteroid gravity levels ($\sim 10^{-4}$ to $10^{-5}g$) for the regolith simulant material SiO$_2$ (expected to have similar cohesive properties to the UCF-1 simulant) occurs for around mm to cm grain sizes. For smaller grains, cohesion dominates, and the target response to a low-velocity impact displays a more elastic behavior, as described in \cite{brisset2018regolith}. Studying mm- to cm-sized grains at the same $g$-levels is useful to understanding the transition between the cohesion- and gravity-dominated regime.

Secondly, recent pictures returned from in-situ space missions \citep[Hayabusa2, ][, and OSIRIS-REx, \citealt{dellagiustina2019properties}]{sugita2019geomorphology} show that the surfaces of small primitive asteroids can be covered in mm- to cm-sized regolith grains, with local depletion of finer grain fractions. As these small bodies have surface gravity levels around 10$^{-5}g$, the granular behavior of this size of regolith grain will dictate the asteroid’s surface morphology and its response to impact or spacecraft contact. Therefore, the study of mm- to cm-sized grains under low levels of gravity has applications for space exploration and the geophysical understanding of small asteroids.

We prepared three types of target material from two grain populations: (1) for a mm-sized population, grains prepared as described above were sieved between 1 and 6~mm grids. As shown in Figure~\ref{f:grains}, the resulting grain size distribution was actually restricted to sizes between 1 and 2~mm, with the distribution peak at 1~mm; (2) for a cm-sized population, the grains remaining on a 1~cm sieve were collected. As shown in Figure~\ref{f:grains}, the resulting grain size distribution was restricted to sizes between 1 and 2~cm, with the distribution peak at 1.3~cm. Figure~\ref{f:grains} also shows pictures of these cm-sized grains, which have an overall angular shape and rough surface; (3) the final one was a mixed target, combining mm- and cm-sized grains at a 1:1 volumetric ratio.

In Table~\ref{t:runs}, we list the target material used for each experiment run: (0) mixed sample; (1) mm-sized sample; and (2) cm-sized sample.

\subsection{Projectiles}
\label{s:projectiles}

The projectiles used in the PRIME-D drops were Teflon, quartz (glass), and brass spheres of 1~cm in diameter. These projectiles were chosen in order to vary the projectile density, and thus the impact energies, over the largest range of values possible while using the same hardware setup. The projectile masses were 4.8, 9.81, and 30.75~g for Teflon, quartz, and brass, respectively. Table~\ref{t:runs} lists which projectiles were used for each experiment run by mass. 

\subsection{Experiment runs}

In Table~\ref{t:runs}, we list the 86 experiment runs we performed using the PRIME-D setup with large grain targets. The experiment plan we followed aimed to include three impacts for each combination of target material (mm- and cm-sized grains), projectile density, and impact velocity range, which was adjusted by varying the launcher spring constant and shim mass. The three impacts for each combination allow for the computation of error bars on the impact outcome. Additional impacts were performed to study the influence on the impact outcome of mixed material and the presence of a target tray.

\begin{figure}[t]
  \begin{center}
  \includegraphics[width = 0.5\textwidth]{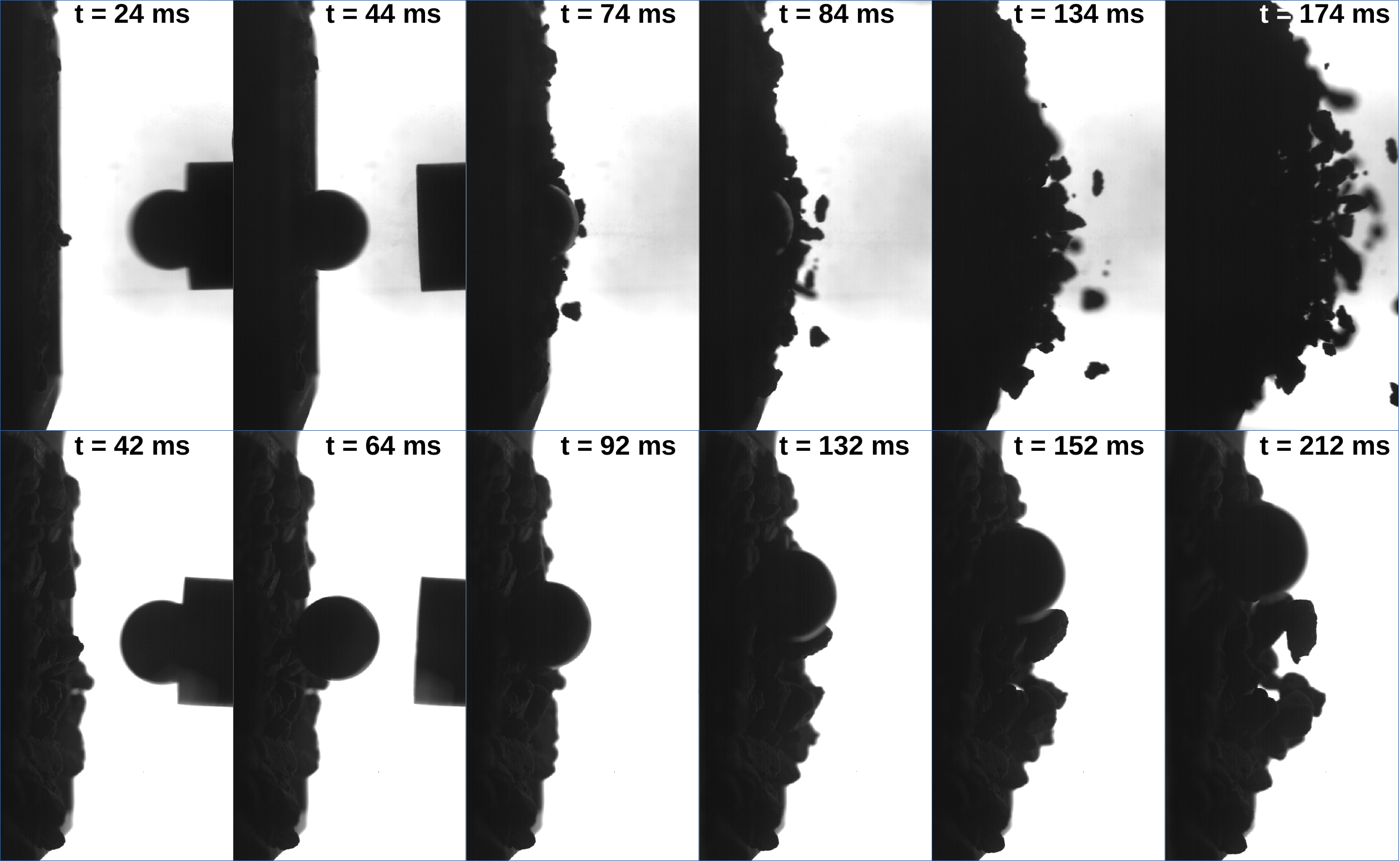}
 \caption{Frame sequences of impacts 2018-02-08-1, brass projectile at 1.3~m/s into mm-sized grains (top) and 2017-08-30-2, glass projectile at 60~cm/s into cm-sized grains (bottom). In these sequences, time elapses from left to right. The launcher is to the right of the image, and the projectile travels toward the target on the left. In early frames, the shim can be seen, which then retracts due to the launcher spring. On the first frames shown, the projectile is already traveling independently of the shim, which can assume angled positions after the launch without impacting the normal trajectory of the projectile.}
 \label{f:impacts}
 \end{center}
\end{figure}

When starting the experiment, we were using the target tray shown in Figure~\ref{f:HW},c to hold the sample. Out of concern for wall effects on the impact outcome for mm- and cm-sized grain targets, we also performed impacts with the target tray removed. For these impacts, we filled the entire bottom half of the experiment vacuum box (see Section~\ref{s:HW}, above) with target material. This provided for a regolith depth of about 40~cm instead of the 3~cm of the tray depth. These impacts are marked at (0) in the "Tray" column of Table~\ref{t:runs}, while the ones making use of the target tray are marked as (1). 

The impact speeds were measured from the high-speed video data collected during experiment runs. After leaving the launcher, the projectile travels at a constant speed on a linear trajectory (impacts are performed in microgravity conditions). In Figure~\ref{f:impacts}, we show characteristic examples of frame sequences obtained during two impacts, one at 1.3~m/s into mm-sized grains generating an ejecta plume (top) and the second at 60~cm/s into cm-sized grains lifting about five target grains (bottom).

Table~\ref{t:runs} also lists the impact outcomes, coefficient of restitution (COR) when projectile rebound took place, and an estimation of the ejected target mass. Details on these characteristics are provided in the following sections.

\begin{figure}[t]
  \begin{center}
  \includegraphics[width = 0.49\textwidth]{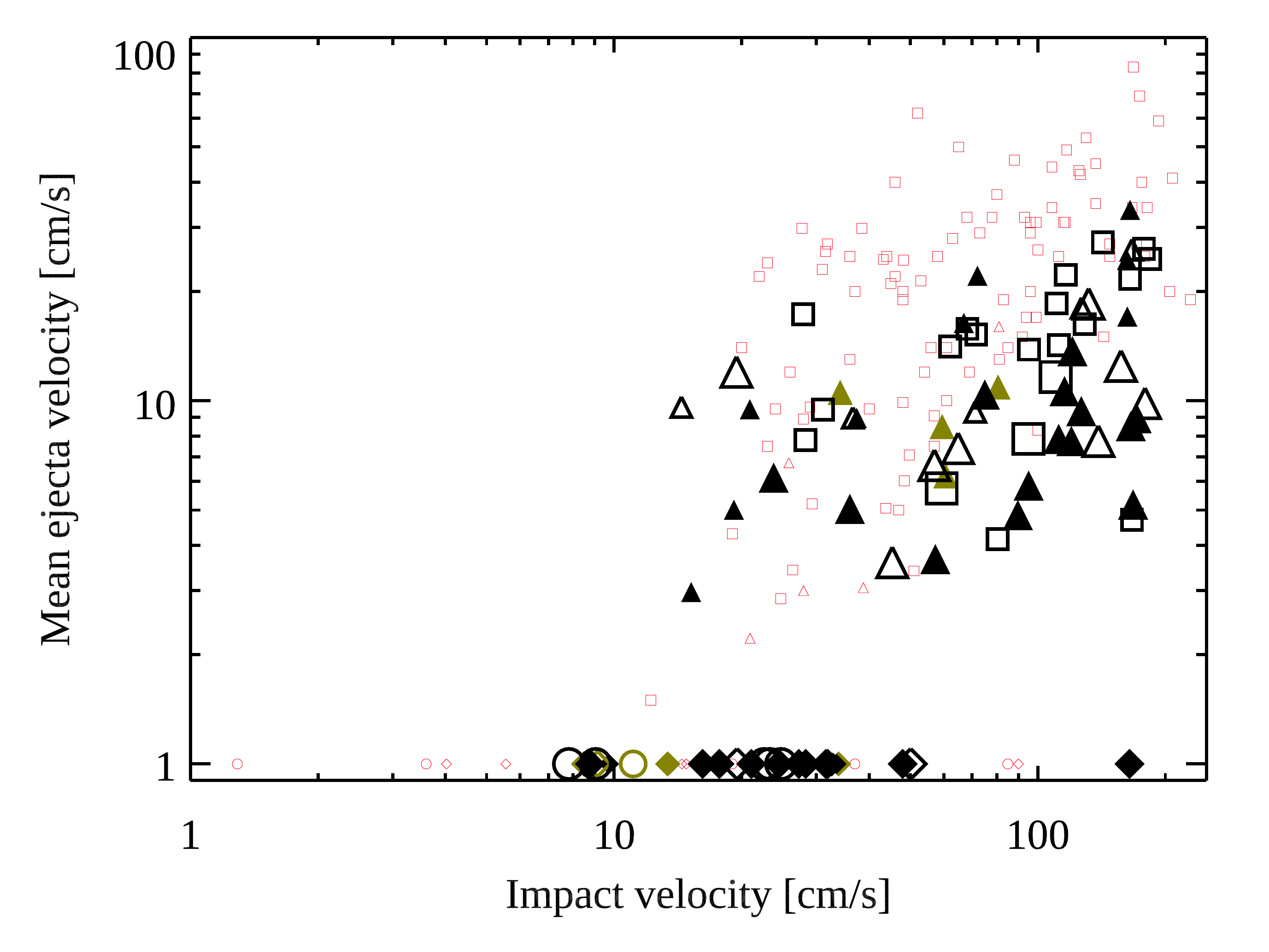}
\begin{picture}(0,0)
\put(-80,150){\includegraphics[height=1.5cm]{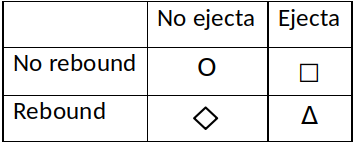}}
\end{picture}
 \caption{Outcome of low-velocity impacts into simulated regolith targets. Circles: projectile embedded, no ejecta. Diamonds: projectile rebound, no ejecta. Squares: projectile embedded, ejecta production. Triangles: projectile rebound, ejecta production. Filled symbols: projectile rolling. Open symbols: no rolling observed. Small symbols: mm-sized grain target. Large symbols: cm-sized grain target. Green symbols: mixed grain target. The impact outcomes for the impacts into fine grains studied in \cite{brisset2018regolith} are shown in red in the background. All impacts producing no ejecta are shown at a mean ejecta velocity of 1 cm/s for the clarity of the graph, even though no speed was measured as no ejecta was observed.}
 \label{f:outcomes}
 \end{center}
\end{figure}

%--------------------------------------------------- 
\section{Data analysis}
\label{s:analysis}

In the following, we describe the data analysis methods we used on our experimental impacts to obtain the results presented in Section~\ref{s:results}. The only data product from the listed experiment runs is a high-speed (500~fps) video of each impact of the projectile into its target. All results described below are sourced from this video data.

\subsection{Classifying impact outcomes and ejecta production}
\label{s:outcomes}

Following the data analysis performed for impacts into fine grains \citep{brisset2018regolith}, we sorted target responses according to ejecta production and projectile rebound. Ejecta production was characterized by an estimation of the ejecta mass produced (last column in Table~\ref{t:runs}): (0) no ejecta; (1) the ejected mass is much lower than the projectile mass; (2) the ejected mass is on the order of the projectile mass; (3) the ejected mass is much higher than the projectile mass. In addition, the trajectories of individual grains ejected from the target after impact were tracked using the open-source ImageJ software. For low-ejecta masses (ejecta mass estimations of 1 or 2), all ejected grains were tracked and their velocities were averaged to obtain a mean ejecta velocity. For high-ejecta masses (an estimation of 3), an ejecta curtain formed and not all grains could be tracked. In that case, we applied the same tracking method as \cite{brisset2018regolith} and tracked 30 ejected grains to obtain a mean ejecta velocity. The resulting ejecta velocities are shown in Figure~\ref{f:outcomes}.

If the projectile did not emerge from the target, the impact outcome was labeled as "embedding" (E in Table~\ref{t:runs}, circles and squares in Figure~\ref{f:outcomes}). If the projectile emerged from the target, the outcome was labeled as "rebound" (R in Table~\ref{t:runs}, diamonds and triangles in Figure~\ref{f:outcomes}).

\begin{figure}[t]
  \begin{center}
  \includegraphics[width = 0.45\textwidth, clip=true, trim={5cm 4cm 12cm, 2cm},angle=90]{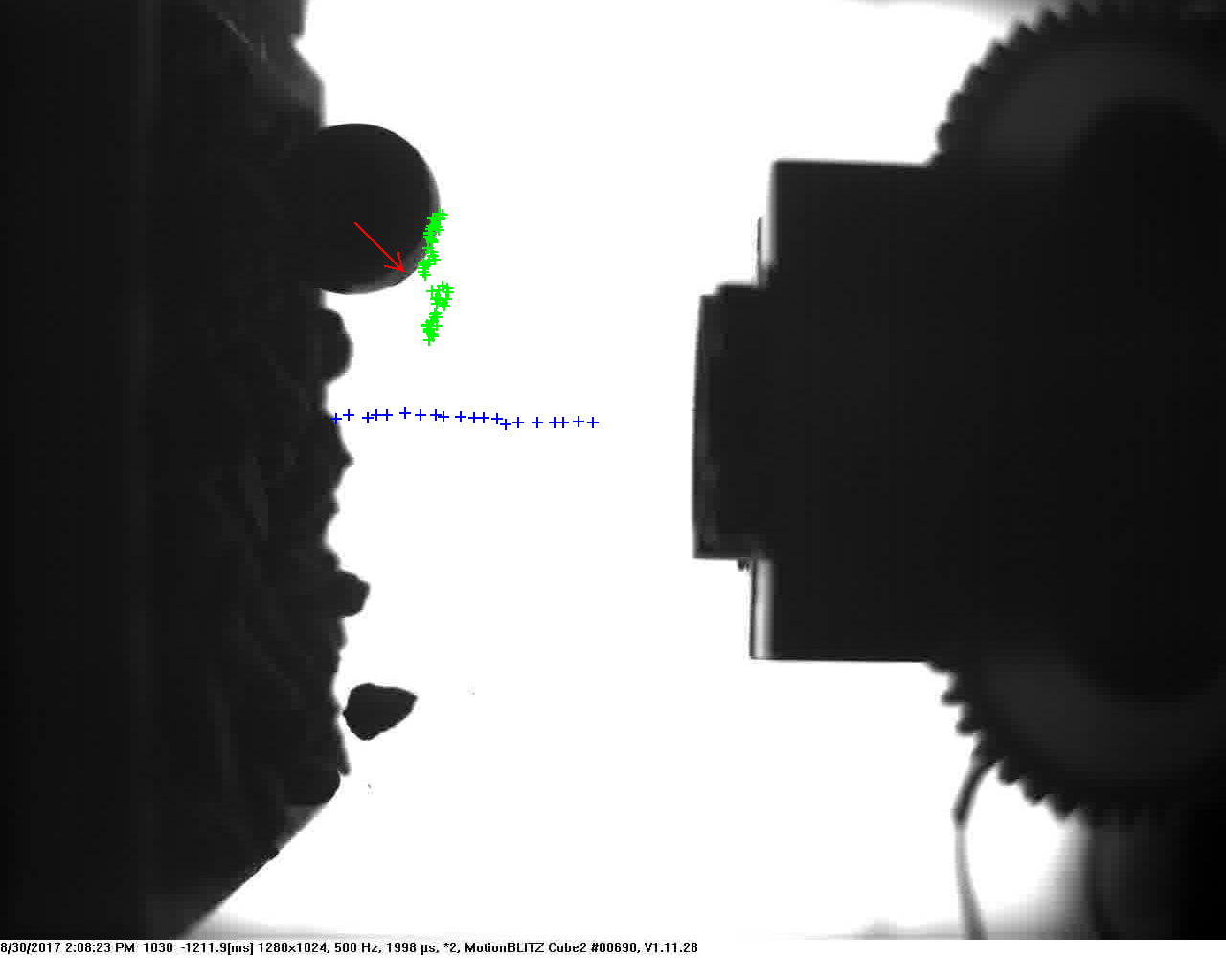}
 \caption{Tracked trajectory of projectile before (blue points) and after (green points) impact for Drop ID 2017-08-30-1. At the top of the image, we can see the launch mechanism that released the projectile in a normal trajectory toward the sample. Before the impact, the bottom of the projectile is tracked to optimally detect contact with the sample, while the top of the projectile is tracked after impact. The projectile motion after impact follows the rough surface of the target material toward the left side of the image. To the right of the image, we can see one ejected particle leaving the target. The red arrow indicates a trackable feature at the surface of the projectile, which was used to verify actual rotation of the projectile during the rolling motion.}
 \label{f:rolling}
 \end{center}
\end{figure}

Compared to fine-grain impacts, we observe an additional outcome for the target response: "rolling" (Ro in Table~\ref{t:runs}, filled symbols in Figure~\ref{f:outcomes}). Rolling is characterized by a sideways motion of the projectile upon impact, following the surface of the target. In Figure~\ref{f:rolling}, we show an example of such a rolling case. Tracks of the projectile show that it roughly follows the surface of the target material. On some of the drop data, we were able to track projectile surface features to confirm actual rotation of the projectile during the rolling motion. Such a feature is shown by a red arrow in Figure~\ref{f:rolling}. The rolling projectile trajectories indicated that the projectile was following the target surface while in intermittent contact with it, as much as a rolling ball would do on irregular ground. We therefore labeled this behavior as rolling. 

\subsection{Projectile rebound}
\label{s:COR}

Projectile rebound was assessed by tracking the projectile trajectory before and after impact using the open-source software ImageJ. Before impact trajectories are characterized by all projectile positions after leaving the launcher and before touching the top edge of the target material. These trajectories were used to calculate the impact speeds shown in Figures \ref{f:outcomes} and \ref{f:espeed_ispeed}, and listed in Table~\ref{t:runs}. If the projectile emerged from the target or rolled on it, we tracked its trajectory after leaving the top edge of the target material and a post-impact velocity was computed. The COR of the impact was calculated using the ratio between the projectile speed after and before impact.

\subsection{Projectile penetration depth}
\label{s:zmax}

In order to study the projectile penetration behavior, we also tracked it during the impact. Tracking of "during-impact" trajectories was started at the moment of first contact between the projectile and the target material. Tracking was stopped when the projectile was either not visible any more (e.g., in the case of an ejecta plume), or when it reached its deepest position within the sample ($z_{max}$). The feature used for tracking was the very top of the spherical projectile in order to ensure accurate tracking for the longest time possible. 

\begin{figure}[t]
  \begin{center}
  \includegraphics[width = 0.49\textwidth]{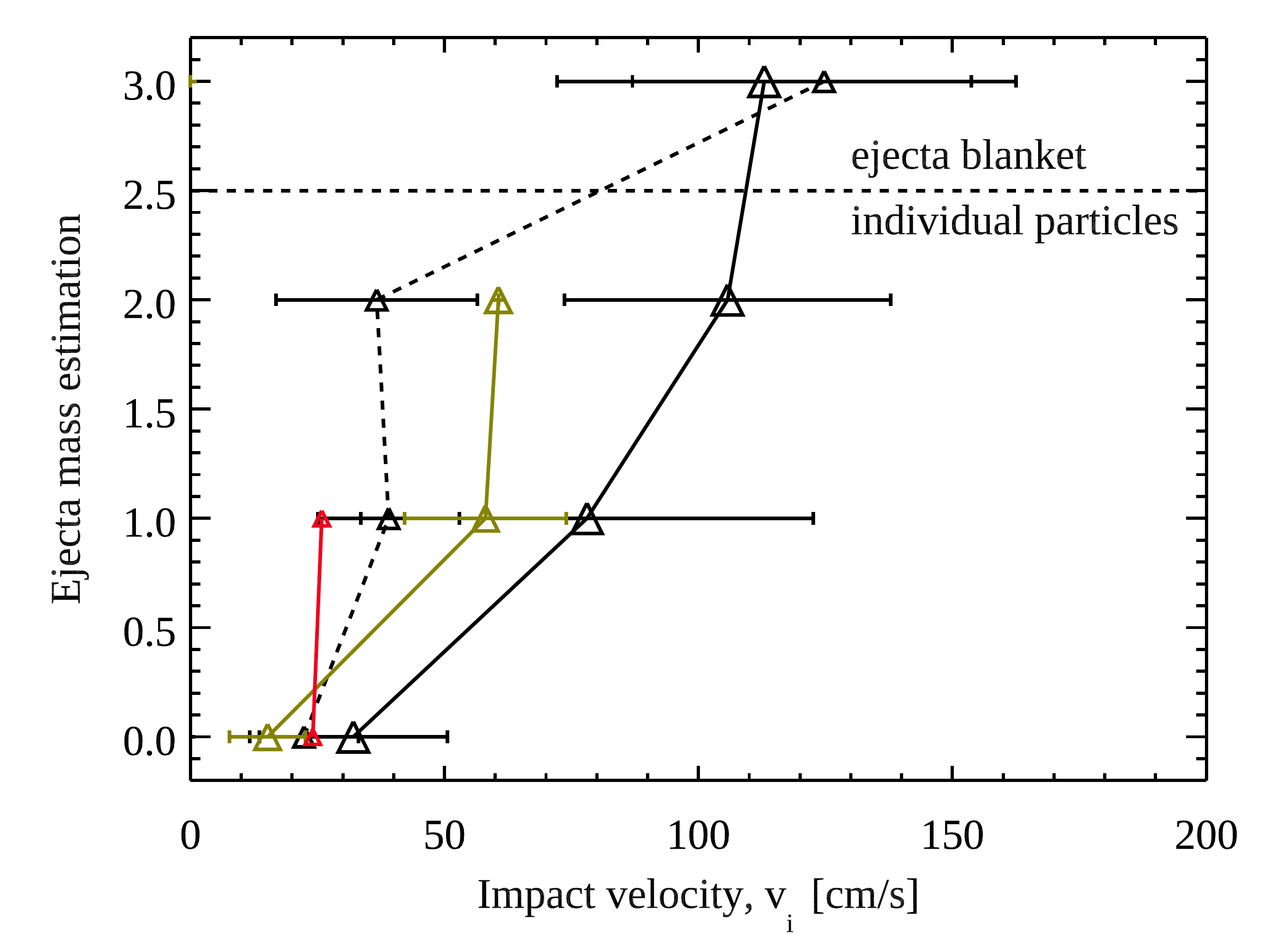}
 \caption{Estimation of ejecta masses for targets composed of 0.25 \citep[full red line, ][]{brisset2018regolith}, 1 (dashed black), and 13~mm grains (full black line). We also show the mass estimations for mixed targets (green line). As described in the text, ejecta production was characterized as follows: (0) no ejecta; (1) the ejected mass is much lower than the projectile mass; (2) the ejected mass is to the order of the projectile mass; (3) the ejected mass is much higher than the projectile mass. For fine grains and mixed targets, only lower ejecta masses were observed. The impact velocity given for each target type is the average of all impacts that created the same amount of ejecta. The horizontal dotted line marks the limit between the ejection of individual particles and an ejecta blanket.}
 \label{f:masses}
 \end{center}
\end{figure}

%------------------------------------------------------------------------------

\section{Results}
\label{s:results}

\begin{figure}[t]
  \begin{center}
  \includegraphics[width = 0.49\textwidth]{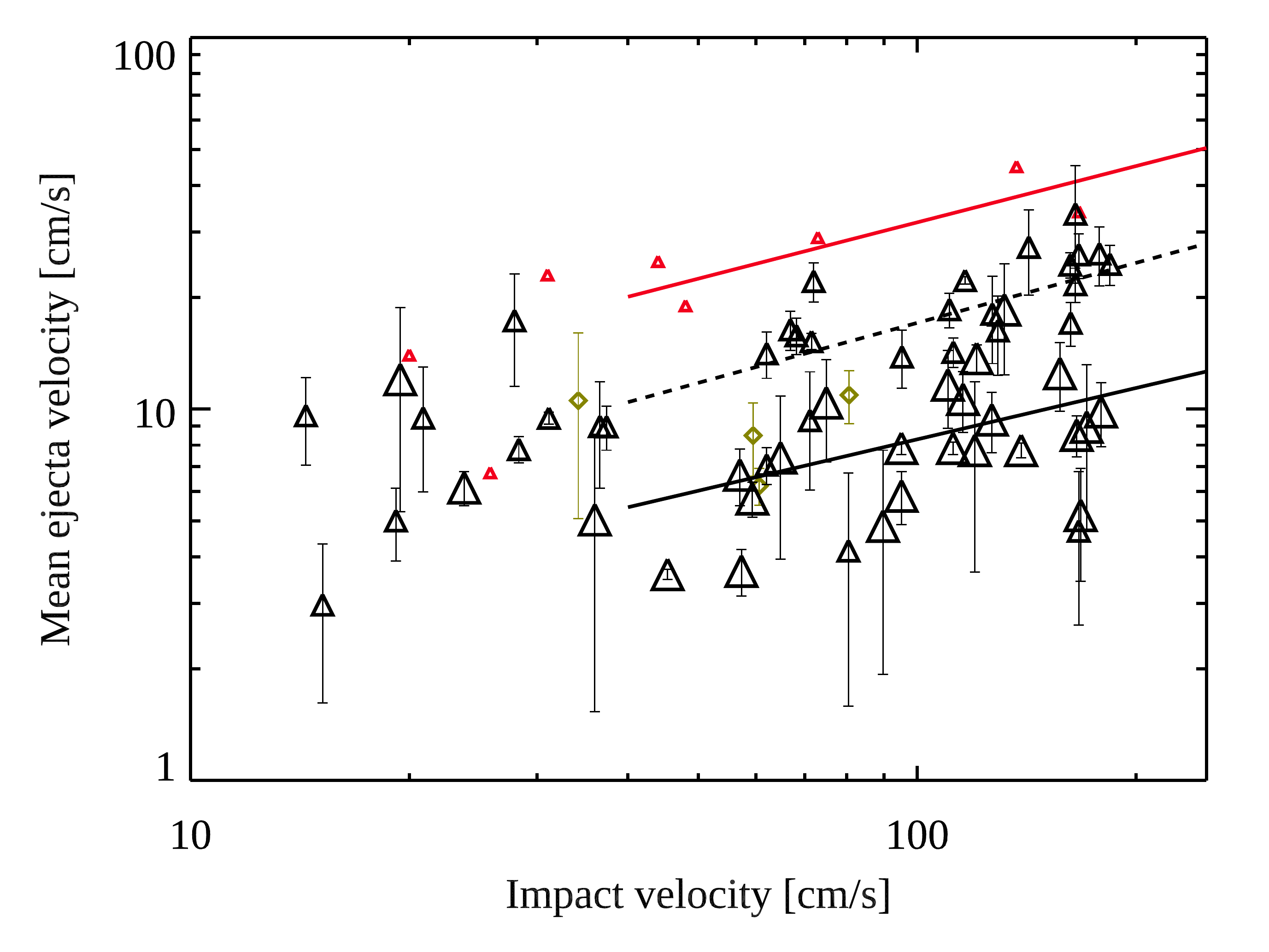}
 \caption{Details of the average ejecta velocity measured for large grain target impacts as a function of the projectile impact velocity. The error bars shown are the mean absolute deviations of ejected grain velocities for each impact. Small symbols: mm-sized grain target. Large symbols: cm-sized grain target. Green symbols: mixed grain target. Red symbols: impacts onto fine JSC-1 grains measured in \cite{colwell2008Icarus} and \cite{brisset2018regolith}. The dashed and dotted lines show the exponential fit for impacts at $>$ 40~cm/s into the mm- (fit index 0.54) and cm-sized targets (index 0.46), respectively. The red line shows the exponential fit determined in \cite{brisset2018regolith} for impacts into JSC-1 fine grain targets, which has an index of 0.50.}
 \label{f:espeed_ispeed}
 \end{center}
\end{figure}

\subsection{Impact outcomes} 

In Figure~\ref{f:masses}, we show the ejecta mass estimations we performed for each impact (see \ref{s:outcomes}) with the average impact velocity at which each outcome was observed. We observe that individual grains separate from mm-sized targets for much lower impact velocities than for cm-sized targets, while an ejecta plume is obtained at, on average, lower velocities for cm-sized targets. In Figure~\ref{f:masses}, we also show the fine grain ejecta mass estimations from \cite{brisset2018regolith} in red and the results for mixed grain targets in green. For the parameter space studied, no ejecta plumes were recorded for these two types of targets. For the lifting of individual particles from the target (mass estimations of 1 and 2), increasing impact velocities are required for increasing grain sizes, as more energy is required to induce motion in larger grains. This trend is reversed for the lifting of ejecta blankets, indicating that the target material cohesion becomes apparent for larger impact energies and that larger grains present a looser surface than smaller ones.

In Figure~\ref{f:outcomes}, we show the outcomes of low-velocity impacts into fine grains studied in \cite{brisset2018regolith} as red symbols. These impacts were generated using on-board flight platforms (parabolic aircraft and suborbital rockets) while the impacts into large grain targets presented here made use of the laboratory drop-tower setup described in Section \ref{s:HW}. While the microgravity quality in the drop tower is comparable to the free-floating aircraft setup producing the lower impact velocities for the fine grain impacts, the available time to observe the impact and following projectile and ejecta behavior is much shorter (0.75~s compared to 20~s on the aircraft). The laboratory drop tower therefore does not allow for impact velocities below about 7~cm/s.

In general, the mean ejecta velocity observed for the impacts on the large grain targets covered the same parts of the parameter space as for impacts into fine grains. Similarly to fine grain targets, ejecta production started for impact speeds between 10 and 20~cm/s. One main difference is that about half the impacts (51~\%) displayed rolling, compared to no occurrences for fine grains.

\begin{figure}[t]
  \begin{center}
  \includegraphics[width = 0.49\textwidth]{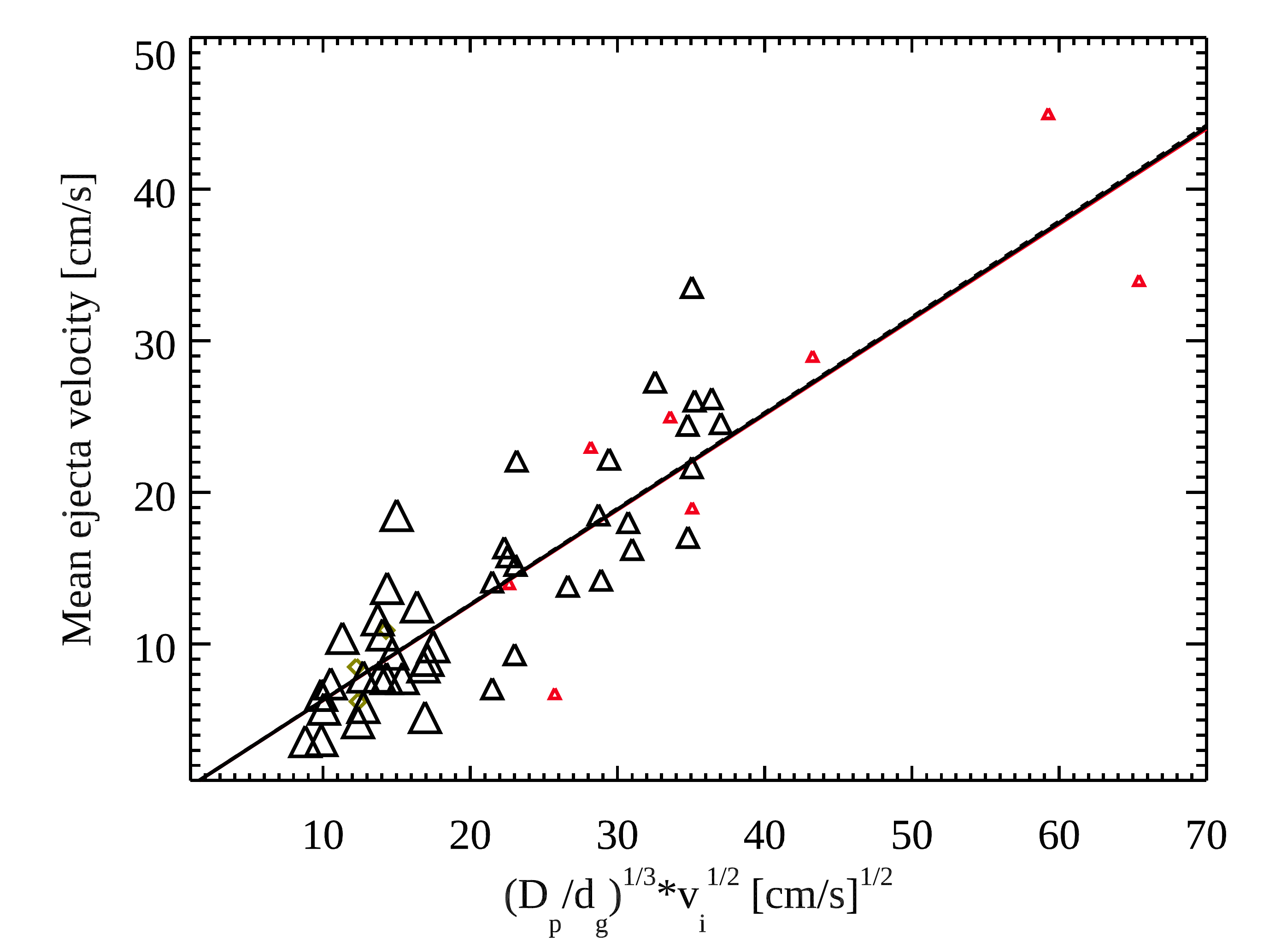}
 \caption{Ejecta speed scaling with projectile diameter ($D_p$), the average target grain diameter ($d_g$), and the impact speed ($v_i$). Symbols and (overlapping) lines are the same as in Figure~\ref{f:espeed_ispeed}. The proportionality factor is 0.63 for all three grain populations.}
 \label{f:espeed_scaling}
 \end{center}
\end{figure}

\subsection{Ejecta speeds}

In Figure~\ref{f:espeed_ispeed}, we show the details of the measured ejecta velocities, including error bars. We note that low impact velocities ($<$ 40~cm/s) lead to similar ejecta velocities in both target types, while smaller grains get ejected faster than larger grains at higher impact velocities ($>$ 40~cm/s). To illustrate this, we show exponential fits for the impacts at speeds above 40~cm/s (dashed and full line for mm- and cm-sized grains, respectively). The associated indices are 0.54 and 0.46, respectively, both close to the 0.50 index obtained for impacts into 0.25~mm JSC-1 grains in \cite{brisset2018regolith}. 

For impact speeds below 40~cm/s, the ejecta speeds for large grain targets are well mixed with results for fine grain targets, with larger ejecta speeds as expected, while coexisting with a large population of impacts not producing any ejecta (Figure~\ref{f:outcomes}). This type of trend at lower impact speeds has been observed in previous data sets \citep{colwell2003Icarus,colwell2008Icarus} and has been interpreted as resulting from the experimental limitations of each set of measurements. In these former data sets, the higher acceleration environment (10$^{-2}g$ rather than 10$^{-4}g$) was considered to be responsible for filtering ejecta to only the fastest grains, keeping the slower ones in the target. In our present experiments, the acceleration environment is below 10$^{-4}g$ and not the cause of the higher ejecta speed bias observed at low impact speed. Instead, it is the short observation time available for each impact. Indeed, only a total of 0.75~s are available for each experiment run in the drop tower (see \ref{s:HW}), which must accommodate for the travel of the impactor to the target and any target response after impact. The slower the impact speed, the longer the projectiles require to travel to the target surface, and the shorter the observation time is available after impact to record ejecta speeds. Therefore, we are able to record only the fastest ejecta particles that emerge from the target first, while slower, later ejecta are missed. This effect biases the data set toward slower impact speeds, with a threshold observed at impact speeds of about 40~cm/s. For this reason, we only consider ejecta produced for impacts at speeds $>$ 40~cm/s in the following analysis.

\begin{figure}[t]
  \begin{center}
  \includegraphics[width = 0.49\textwidth]{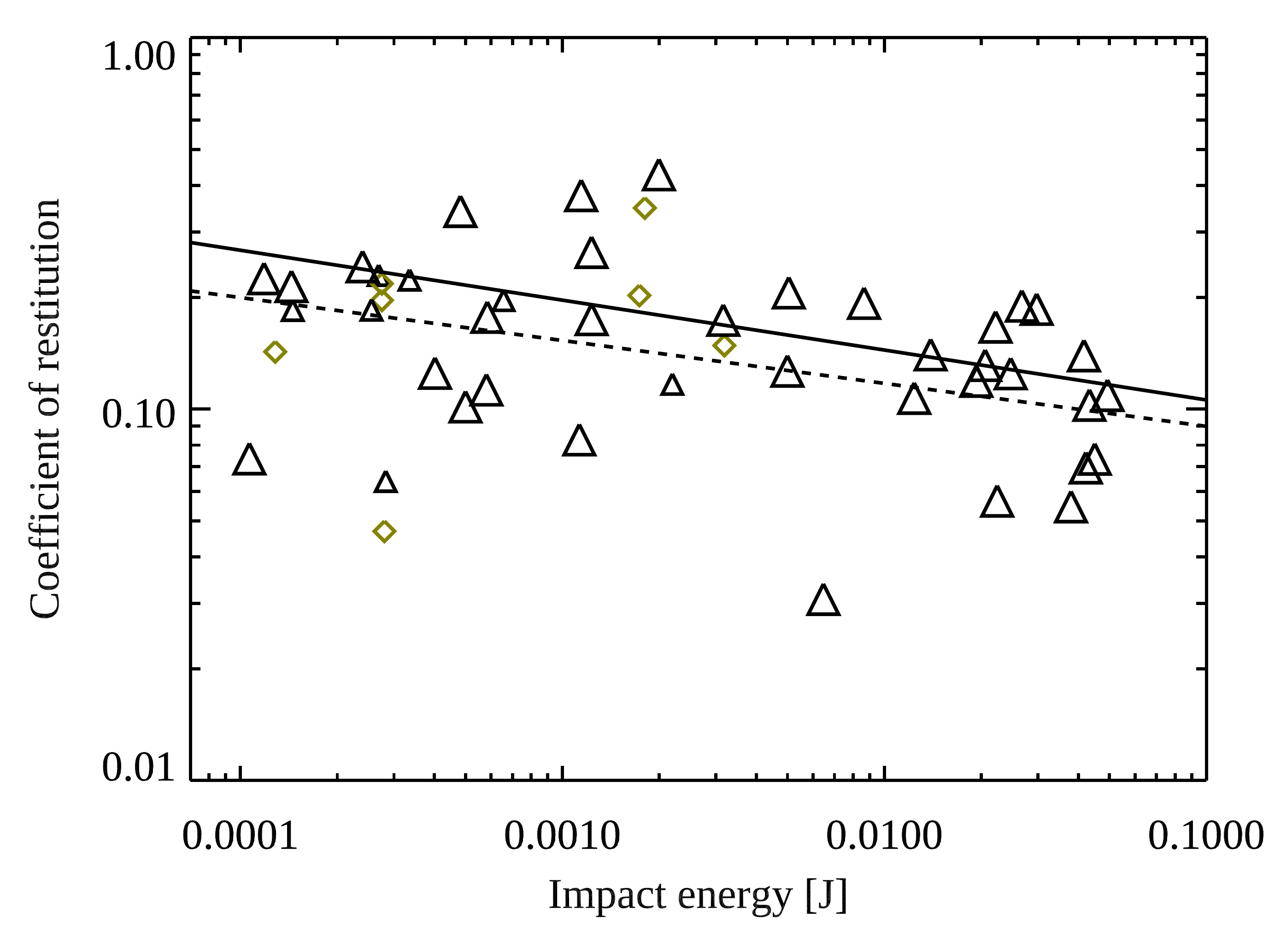}
 \caption{Total coefficient of restitution for projectiles launched onto large grain targets and rebounding or rolling after impact. Symbols and lines are the same as in Figure~\ref{f:espeed_ispeed}. Indices for the exponential fits are -0.12 and -0.13 for the mm- and cm-sized grain targets, respectively.}
 \label{f:COR}
 \end{center}
\end{figure}

In Figure~\ref{f:espeed_ispeed}, the red line shows the exponential fit determined for ejecta speed from impacts into fine (0.25~mm) JSC-1 grains in \cite{brisset2018regolith}. We notice that ejecta from all three target types, fine, mm-, and cm-sized grains, seem to scale in a similar manner with the projectile impact speed, that is, an index of 0.5 with an offset for each grain size. We attempted a scaling of the ejecta speeds with the grain size, by introducing the dimensionless quantity $D_p/d_g$, $D_p$ being the projectile diameter and $d_g$ the average target grain diameter. For the impacts shown in Figure~\ref{f:espeed_ispeed} and used for the scaling, the projectile diameter was either 1.9~cm \citep[fine grains, see ][ and \citealt{brisset2018regolith}]{colwell2008Icarus} or 2~cm (large grains, see \ref{s:projectiles}). In Figure~\ref{f:espeed_scaling}, we show the result of our ejecta speed scaling. We were able to find a proportionality relationship between all measured ejecta speeds and the quantity $(D_p/d_g)^{1/3}v_i^{1/2}$, with identical proportionality factors to the second decimal for the three grain populations (0.63). While the linear fits overlap very well, we observe a non-negligible remaining spread in the data. This could be due to a number of factors, including the width of the distribution in the grain size in each target.

Only three data points for impact speeds above 40~cm/s did not allow for proper data fitting, but when applying a target grain size of 5~mm for these impacts (averaged between the mm- and cm-sized populations), we find that these data points collapse within the data spread. This indicates that mixed targets also follow the proposed scaling. 

\subsection{Projectile Rebound}
\label{s:COR2}

In Figure~\ref{f:COR}, we show the measured CORs for rebound and rolling impacts into large grain targets versus the impact energy of the projectile. The indices of the exponential fits are -0.12 and -0.13 for mm- and cm-sized targets, respectively, while it is at 0.07 for mixed grain targets. This shows an COR increase for lower impact velocities for mm- and cm-sized grain targets and practically no dependance on the impact velocity for mixed targets.

Average values are similar at 0.16 $\pm$ 0.07 and 0.17 $\pm$ 0.05 for mm- and cm-sized targets, respectively. Mixed grain targets produced an average COR of 0.21 $\pm$ 0.05, which is within the error ranges of the monodisperse targets.

The COR average values found for our impacts into large grain targets were similar for mm- and cm-sized grains at 0.16 and 0.17, respectively (see Section \ref{s:COR}). These values are in good agreement with measurements taken by \cite{brisset2018regolith}, where the COR average for impacts in the same speed range into fine grain target (under microgravity conditions: free-floating box on-board the parabolic aircraft), was found to be 0.15 $\pm$ 0.04. This indicates that the COR of a low-speed projectile rebounding a regolith surface in microgravity is not sensitive to the regolith particle size.

\begin{figure}[t]
  \begin{center}
  \includegraphics[width = 0.49\textwidth]{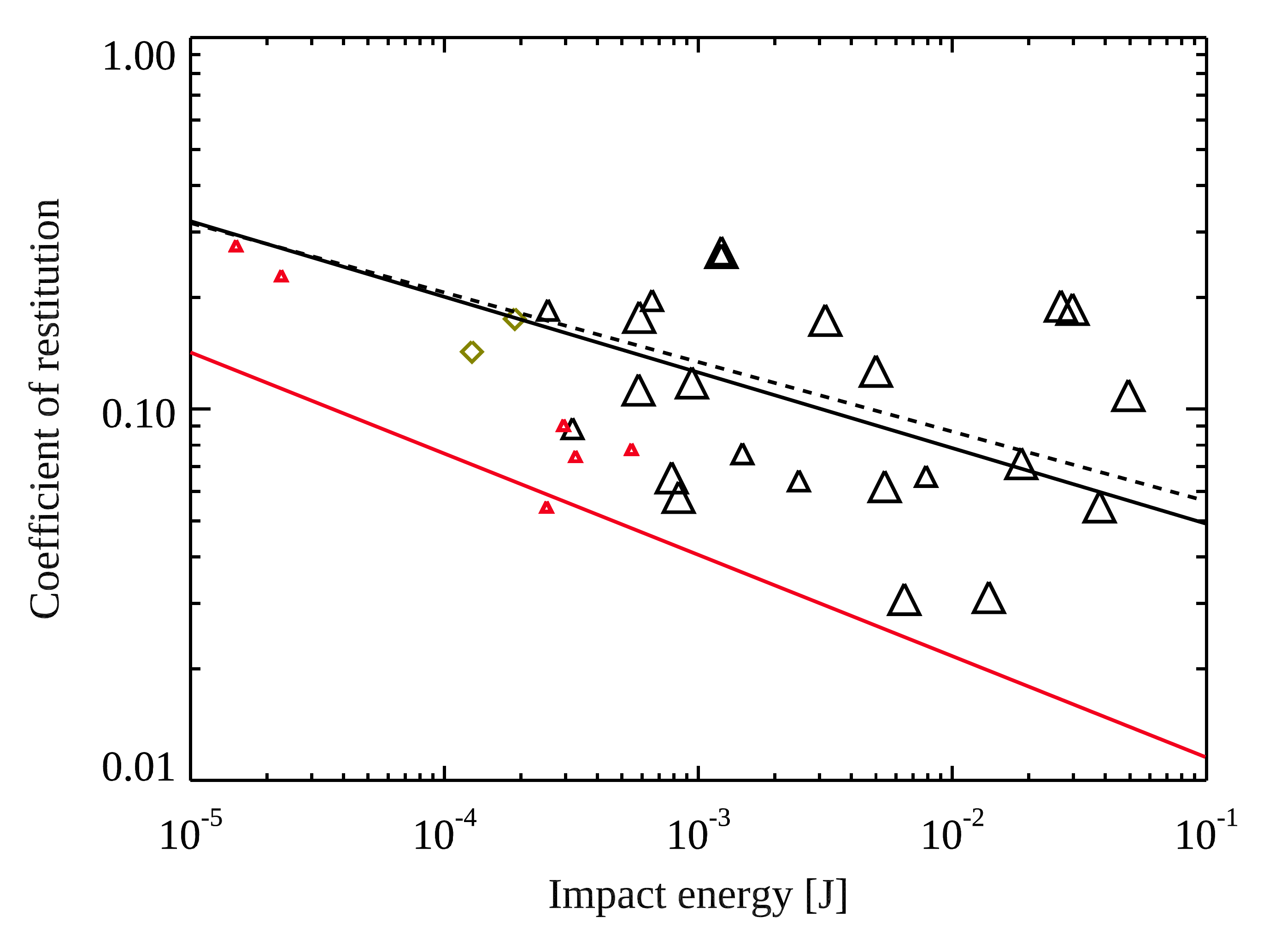}
 \caption{Coefficient of restitution for normal rebound impact outcomes, neglecting rolling outcomes. Symbols and lines are the same as in Figure~\ref{f:COR}. The exponential fit indices are -0.19 and -0.20 for mm- and cm-sized grain targets, respectively. The red line shows the COR fit for impacts into fine grain targets \citep{brisset2018regolith}. Red data points show individual COR values for these impacts, selected for JSC-1 targets and microgravity condition only.}
 \label{f:COR_norolling}
 \end{center}
\end{figure}

In Figure~\ref{f:COR_norolling}, the red line shows the fine grain fit presented in \cite{brisset2018regolith}, which had an index of -0.27 and included data from four flight campaigns. We note that the coefficient of restitution on large grain surfaces is higher on average by about a factor of 2. However, if we only select data points that were collected using a JSC-1 target under microgravity conditions \citep[red points in Figure~\ref{f:COR_norolling}, see ][]{brisset2018regolith}, we can see that these are within the data spread of the large grain data with a convergence towards very low impact speeds. This supports the idea of a low dependence of the COR on the target grain size.

We also note that the CORs recorded in fine grains under 1$g$ conditions range from 0.2 at 10$^{-5}$~J to 0.06 at 10$^{-3}$~J, with an exponential fit index of -0.25 \citep{katsuragi2017}. This is also within our data spread, so that it seems the measurement method employed in \cite{katsuragi2017} allows for a decent approximation of fine grain target behavior in microgravity.

\subsection{Projectile rolling}
\label{s:rolling2}

\begin{figure}[t]
  \begin{center}
  \includegraphics[width = 0.49\textwidth]{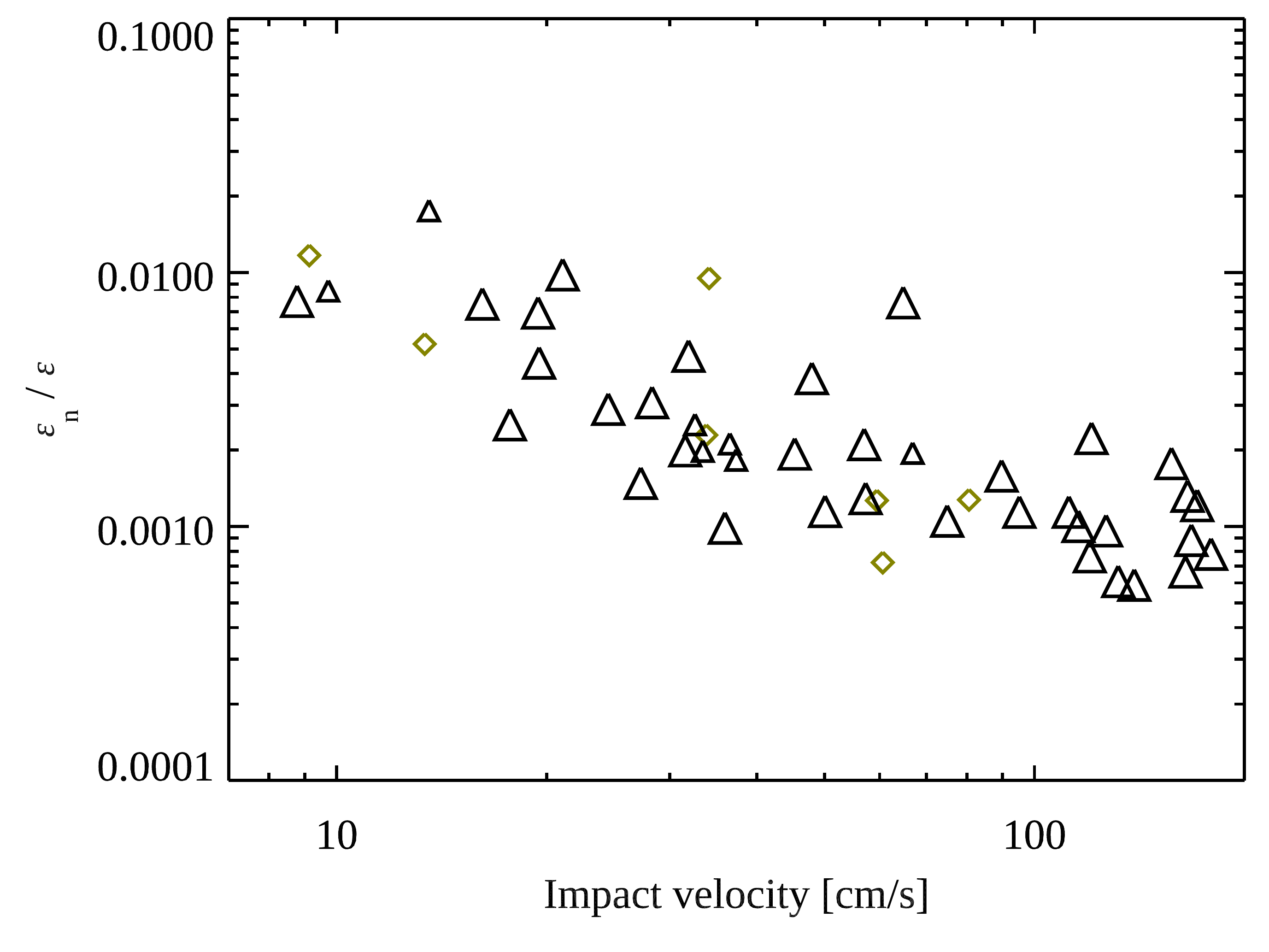}
 \caption{Ratio between normal ($\epsilon_n$) and total ($\epsilon$) coefficient of restitution for impacts displaying rolling as an outcome (see Table~\ref{t:runs}). Symbols are the same as in Figure~\ref{f:espeed_ispeed}.}
 \label{f:COR_ratio}
 \end{center}
\end{figure}

About half the impacts listed in Table~\ref{t:runs} (51~\%) displayed rolling. In an attempt to quantify the amount of sideways motion of the projectile after impact, we calculated the ratio between the normal and total coefficient of restitution for impacts resulting in rolling (Figure~\ref{f:COR_ratio}). The normal coefficient of restitution is defined by the ratio between the normal component of the rebound velocity over the impact velocity of the projectile.

In Figure~\ref{f:COR_ratio}, higher ratios indicate that more of the restituted impact energy remains in the same motion direction as before impact (normal), while low ratios show that most of the restituted energy went into motion in the direction perpendicular to normal (tangential). We can see that the conversion from normal to tangential motion is very efficient ($>$ 99~\% of the restituted energy is converted) and more pronounced at higher impact speeds. For impacts above about 1~m/s, rolling was observed only in cm-sized targets (mm-sized targets have a tendency to generate ejecta plumes at these speeds).

We attribute the occurrence of rolling in the impacts presented here to the reduced size ratio between the projectile and the target grains. Indeed, during impacts into grains with sizes 100 times smaller than the projectile, the projectile interacts with a great number of grains, and the contact surface with the target approximately corresponds to the projectile surface. In the impacts presented here, target grains have sizes only ten times smaller than the projectile (mm-sized grains), or even to the same order (cm-sized grains). In such configurations, the interfacing surface between a projectile and target can be significantly reduced and more importantly, non-normal, despite a normal impact trajectory. For example, the incoming projectile can first come into contact with the target in the form of a single cm-sized grain, of which the exposed surface is at 45$\degree$ with the surface normal. In that case, the rebound on the target surface will send the projectile on a trajectory that is perpendicular to the normal incident, which we observe as rolling in our data. This idea is supported by the observations made in \cite{guttler2012cratering} and \cite{barnouin2019impacts} that small differences in the location of first contact between a projectile and a coarse-grained target tend to play an important role in the impact outcome (such as crater size and shape in the case of their higher-speed impacts).

Given the amount of impact energy that is transferred to tangential motion, and how this outcome is different from the normal rebound we observed on fine grain targets \citep{brisset2018regolith}, we revisited our COR data presented in Figure~\ref{f:COR}, selecting the normal rebounds (R in Table~\ref{t:runs}) and leaving out the rolling (Ro) data points. In Figure~\ref{f:COR_norolling}, we show the result of this data reduction: mm- and cm-sized targets have almost identical exponential fits at 0.04E$^{-0.19}$ (dashed line) and 0.03E$^{-0.20}$ (full line), respectively, with $E$ being the impact energy. We conclude that the difference in COR shown in Figure~\ref{f:COR} between mm- and cm-sized targets seems to be generated mostly by rolling impacts, and both target types behave similarly for normal rebounds.

\subsection{Projectile penetration depth}
\label{s:penetration_depth}

\begin{figure}[t]
  \begin{center}
  \includegraphics[width = 0.49\textwidth]{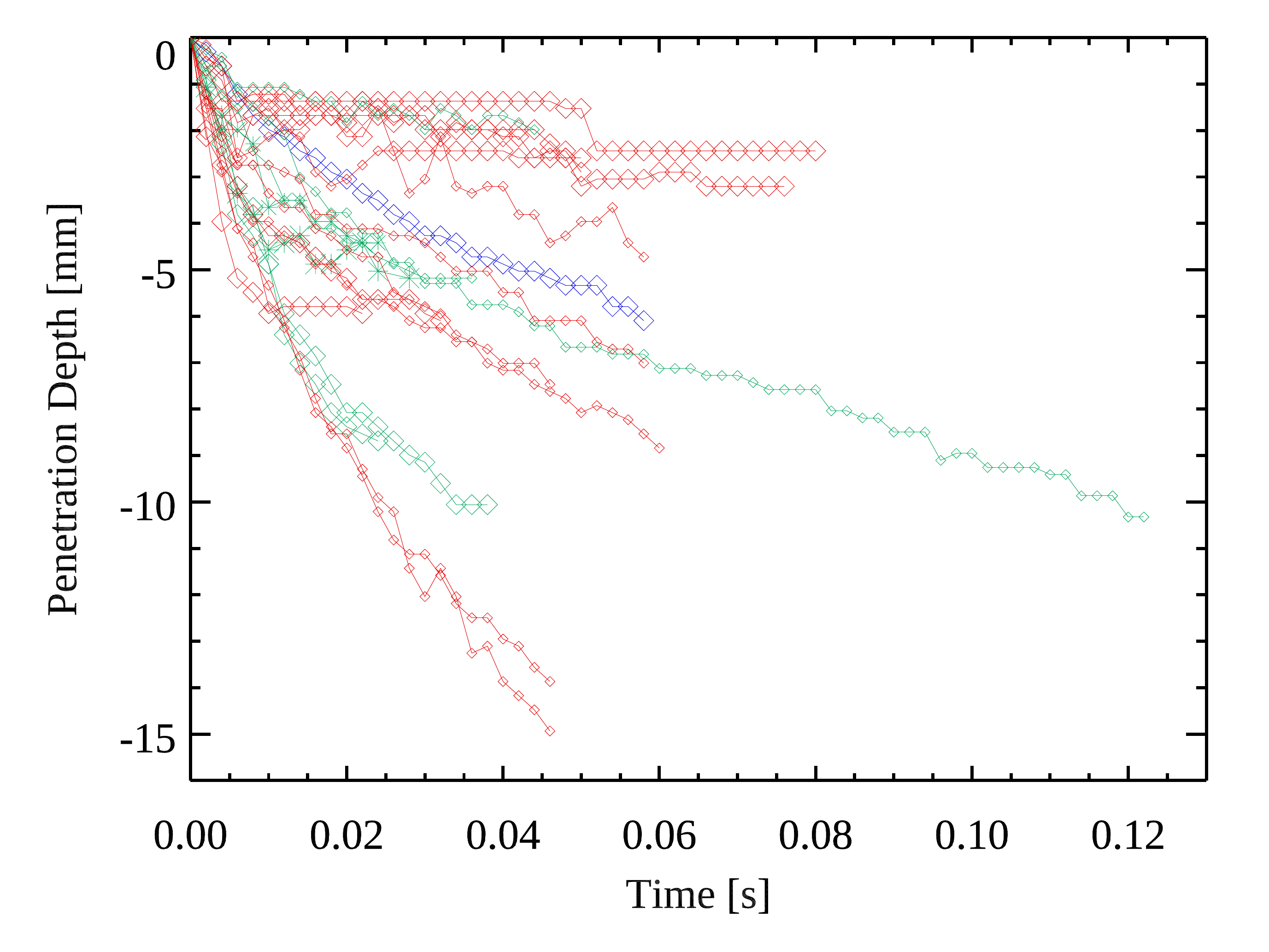}
 \caption{Sample of tracked projectile trajectories during target penetration. Plus signs: Teflon projectile. Asterisk: glass. Diamonds: brass. The symbol sizes indicate the grain size of the target material, with smaller symbols for the mm-sized target grains, medium ones for mixed targets, and larger symbols for cm-sized targets. The color denotes ranges in impact energy, with blue for impacts below 10$^{-3}$~J, green between 10$^{-3}$~J and 10$^{-2}$~J, and red above 10$^{-2}$~J.}
 \label{f:penetration_tracks}
 \end{center}
\end{figure}

\subsubsection{Penetration trajectories} 
In Figure~\ref{f:penetration_tracks}, we show a sample of tracked trajectories during target penetration. We note that the longest and deepest trajectories occurred for brass projectiles, so Teflon and glass projectile trajectories are barely visible in the presented sample. In addition, the deepest penetration occured for impacts at higher energies (dark blue) into mm-sized targets (small symbols), while the shallowest penetrations occured for the same energy ranges, but in cm-sized targets (large symbols). A simple relationship between impact energy and penetration depth is not observed (light blue trajectories are found between dark blue ones). 

\begin{figure}[t]
  \begin{center}
  \includegraphics[width = 0.49\textwidth]{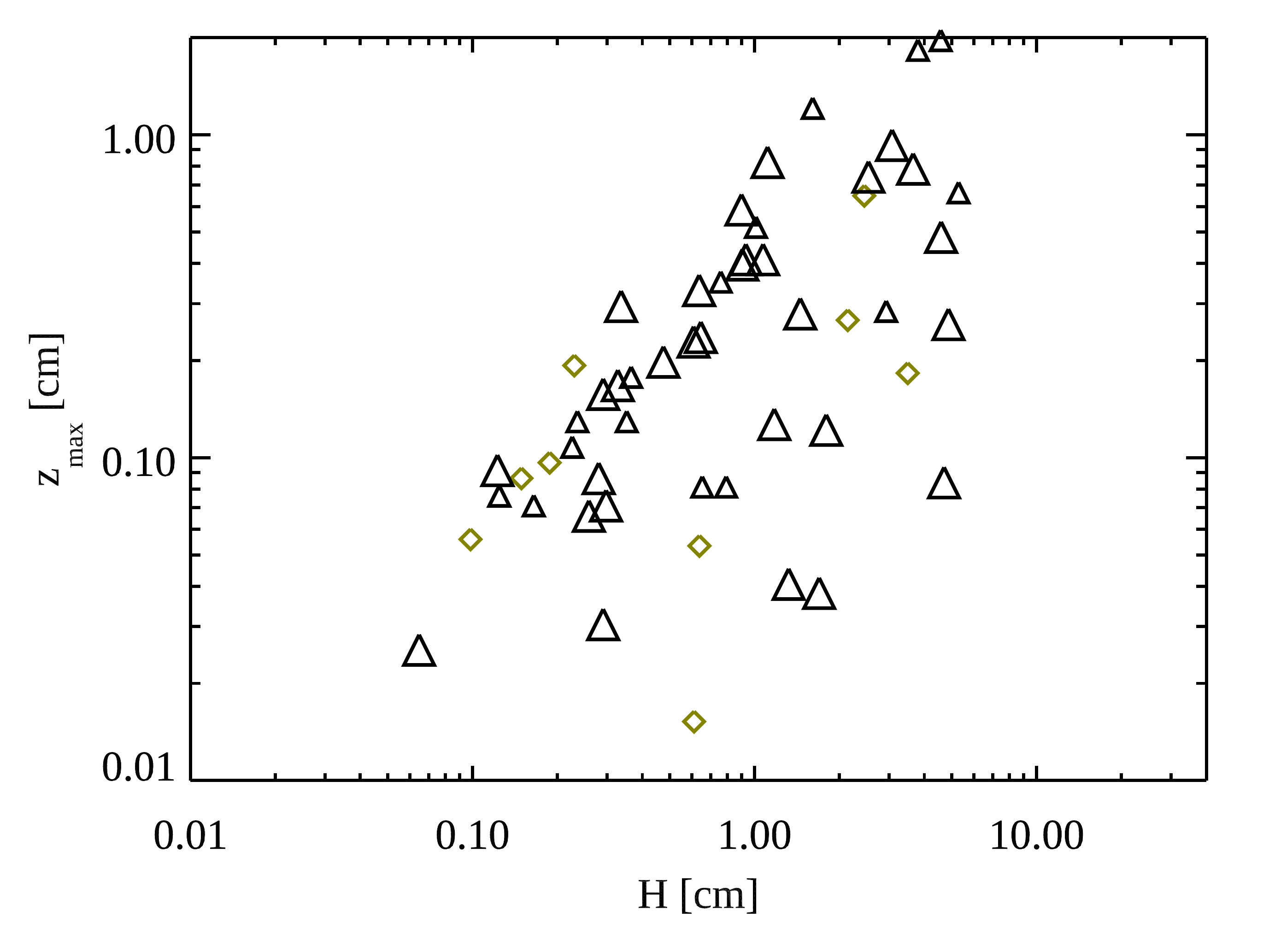}
 \caption{Maximum penetration depth of projectile in target material as a function of the projectile travel height $H$ (see text). Symbols are the same as in Figure~\ref{f:espeed_ispeed}.}
 \label{f:zmax}
 \end{center}
\end{figure}

\subsubsection{Maximum penetration depth} 
From the trajectories during impact, we measured the deepest projectile penetration depth. Results are shown in Figure~\ref{f:zmax}. Following \cite{brisset2018regolith}, we calculated the equivalent projectile travel height $H$ in order to compare our results with 1$g$ experiments presented in \cite{katsuragi2017}: $H = h + z_{max}$, with $h = \frac{v_i^2}{19.62}$ [cm] the 1$g$ equivalent drop height of the projectile \citep[a measure of the impact energy, see][]{brisset2018regolith}, and $z_{max}$ the maximum penetration depth of the projectile into the target. In Figure~\ref{f:zmax}, we notice a concentration of data points around the identity line, indicating low impact velocities ($h \sim 0$). Exponential fits to data from the three target types yield indices of 0.35, 0.72, and 0.51 for cm-size, mm-sized, and mixed grains, respectively. For 1$g$ measurements performed on fine grain targets ($\sim$ 100 $\mu$m grains), \cite{katsuragi2017} found $z_{max} \propto H^{1/3}$, which corresponds to a scaling by impact energy. However, \cite{guettler_et_al2009ApJ} find a better match with momentum scaling with 1$g$ drops into fluffy aggregates composed of 1~$\mu$m-sized grains. Similar measurements under microgravity conditions performed by \cite{brisset2018regolith} showed yet another scaling ($\propto H$). These variations in scaling between 1$g$ and microgravity measurements may relate to the differences in energy dissipation inside the sample. Below we refine the scaling presented in \cite{brisset2018regolith} to include grain size.

\subsubsection{Scaling the penetration depth} 
In \cite{brisset2018regolith}, we were able to scale the maximum penetration depth ($z_{max}$) of the projectile into fine grain targets using the quantity $\mu^{-1}(\rho_p/\rho_g)^{0.1}D_p^{2/3}H^{1/3}$, where $\mu$ is the target material friction coefficient (the tangent of the angle of repose), $\rho_p$ and $\rho_g$ the projectile and target densities, $D_p$ the projectile diameter, and $H$ the equivalent projectile travel height (this quantity is a measure of the impact energy, see Section~\ref{s:zmax} for details). For these fine grain experiments, we found that $z_{max}$ was following an exponential scaling with this quantity with an index of 2.5.

\begin{figure*}[th]
    \centering
    \begin{minipage}{0.5\textwidth}
        \centering
        \includegraphics[width=0.88\textwidth]{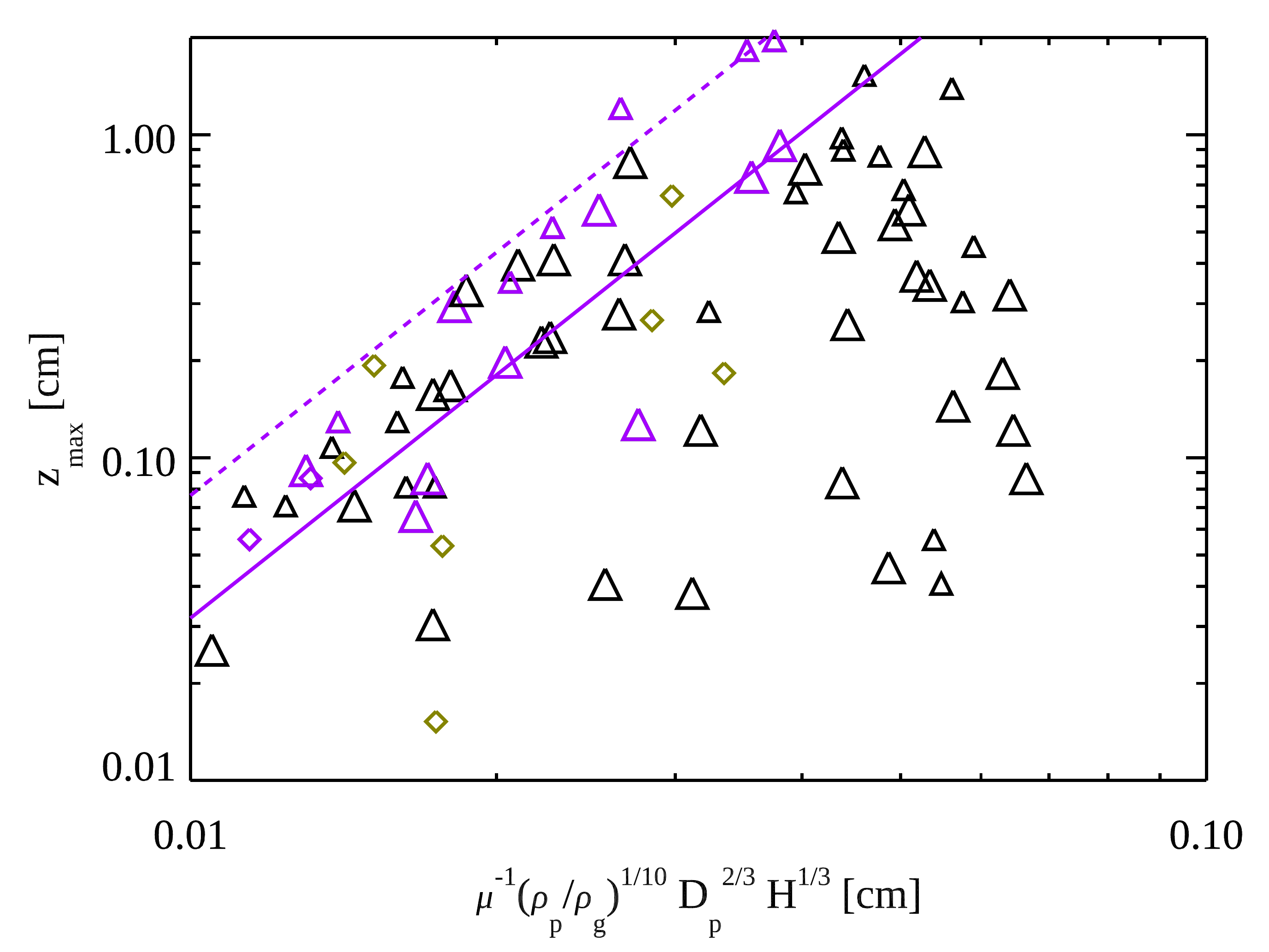} 
    \end{minipage}\hfill
    \begin{minipage}{0.5\textwidth}
        \centering
        \includegraphics[width=0.88\textwidth]{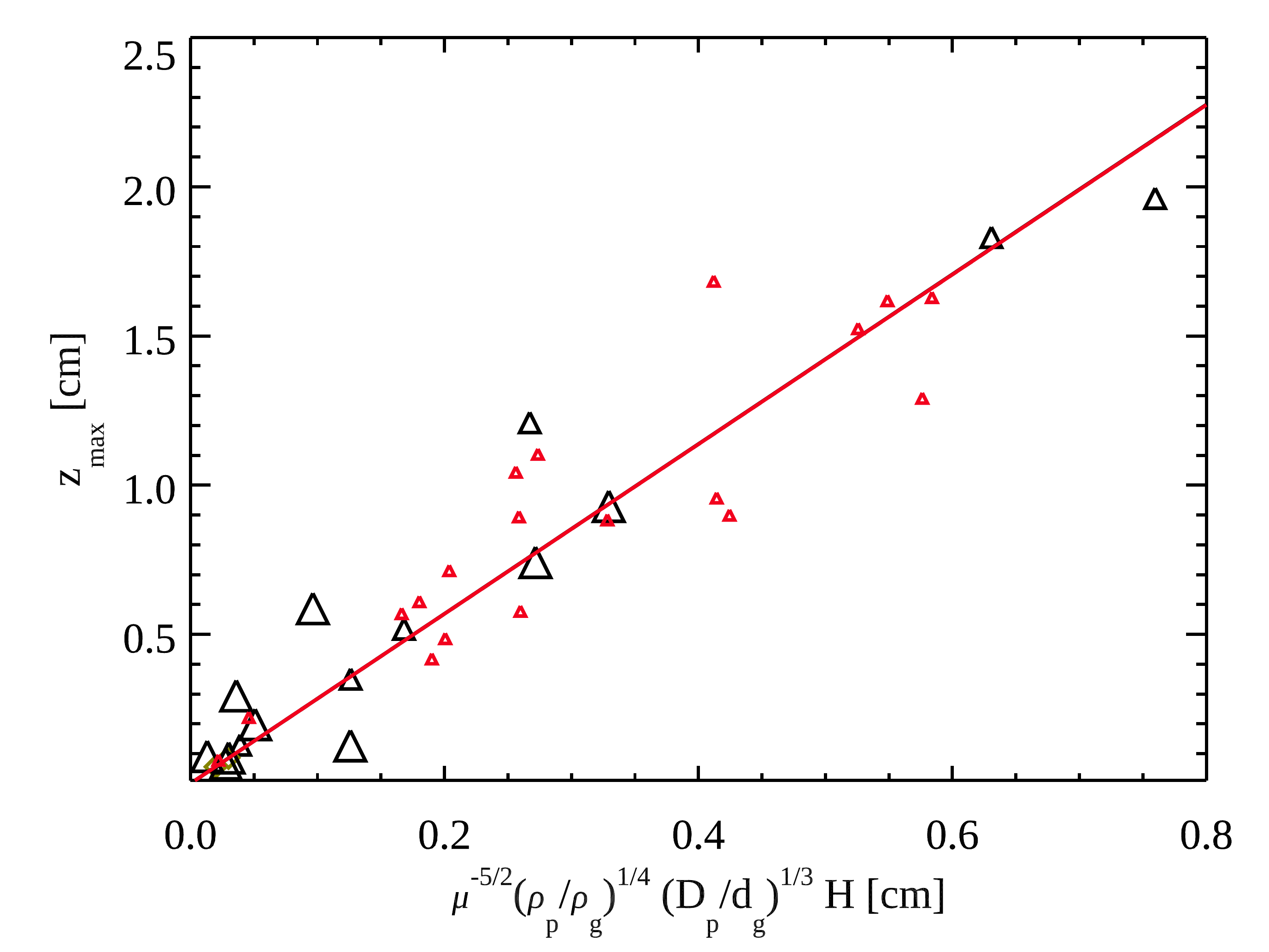} 
    \end{minipage}
    \caption{Maximum penetration depth scaling: (left) maximum penetration depth in dependence of the scaling quantity $\mu^{-1}(\rho_p/\rho_g)^{0.1}D_p^{2/3}H^{1/3}$ (see text for details). Symbol size and color are the same as in Figure~\ref{f:COR}. Purple symbols denote impacts for which rebound was observed without rolling, and purple lines are exponential fits to this subset of data. The fit indices are 2.3 (dashed) and 2.6 (full line) for mm- and cm-sized grain targets, respectively; (right) maximum penetration depth scaling adjusted for inclusion of the target grain size. Red symbols show the fine grain data points collected in \cite{brisset2018regolith}. The three data sets (0.25, 1, and 13~mm grains) are proportional to the quantity $\mu^{-5/2}(\rho_p/\rho_g)^{1/4}(D_p/d_g)^{1/3}H$ with a factor of 2.84.}
    \label{f:zmax_scaling}
\end{figure*}

In order to attempt the same scaling for large grain targets, we measured the relevant quantities for the three target types and logged them into Table~\ref{t:scaling_parameters}. The projectile diameter is 2~cm for all projectiles, and the densities are 1.15, 2.34, and 7.34 g/cm$^3$ for Teflon, glass, and brass projectiles, respectively. In the left section of  Figure~\ref{f:zmax_scaling}, we show the dependency of $z_{max}$ on the calculated scaling quantity for the impacts listed in Table~\ref{t:runs}. No obvious trend can be noticed. 

\begin{table}[b]
\caption{Measured sample properties}
\begin{tabular}{|l|l|l|l|}
\hline
\textbf{Parameter} & \textbf{Mm-sized} & \textbf{Cm-sized} & \textbf{Mixed} \\ \hline
$\mu$ & 0.57 & 0.49 & 0.43 \\ \hline
$\rho_g$ & 1.01 g/cm$^3$ & 0.96 g/cm$^3$ & 0.98 g/cm$^3$ \\ \hline
\end{tabular}
\label{t:scaling_parameters}
\end{table}

However, this scaling was previously applied to normal impacts into fine grains, for which the projectile was entering the target normally after first contact with the target surface. As described in Section~\ref{s:rolling2}, half of the impacts we observed resulted in the projectile rolling sideways on the surface of the target. In these cases, we can expect the target penetration to have a tangential component in addition to the normal one. This would result in shorter penetration depths as part of the motion is deviated sideways, explaining the larger spread of data in Figure~\ref{f:zmax_scaling}, left.

Therefore, we only kept the impacts for which no rolling was observed (marked in purple in Figure~\ref{f:zmax_scaling}, left. Exponential fits to the data for mm- and cm-sized targets yield indices of 2.3 (dashed) and 2.6 (full purple line), respectively, which are both very close to the scaling index of 2.5 found for fine grain targets. This indicates that the scaling defined in \cite{brisset2018regolith} for fine grain targets in microgravity can also be applied to larger grain targets, with a required adaption to include the target grain size. 

In the right-hand side of Figure~\ref{f:zmax_scaling}, we collapse the three data sets for fine grains (0.25~mm diameter, red symbols and line), mm-sized grains (small black symbols, dashed line), and cm-sized grains (large symbols, full line), by including the target grain diameter $d_g$ in the scaling quantity as part of the dimensionless ratio $D_p/d_g$. In order to collapse the three data sets, this ratio had to be included with an exponent of 2/15. After replacing $D_p$ with a dimensionless parameter in our scaling expression, we are left with a quantity of the unit given in cm$^{1/3}$. Given the exponent of the scaling of 2.5 (5/2), we attempted to scale the penetration depth with the very close quantity of $H^{2/5}$ rather than $H^{1/3}$ in order to obtain a final scaling result in cm. The exponential fits to our three data sets collapse to the second decimal for the parameters of the studied impacts (2.84).

We note that only two impacts into mixed grains did not display any rolling of the projectile after impact. We therefore did not perform an exponential fit for mixed grains. However, applying a grain diameter of 5~mm (an average between 1 and 10~mm), we observe that these two impacts also follow the scaling presented in Figure~\ref{f:zmax_scaling}, with an offset. These two data points also collapse to the proposed scaling to the second decimal if the average grain size used in the calculation is 1.1~mm rather than 5~mm. This could indicate that the behavior of mm-sized grains dominates the behavior of the mixed sample when it comes to the penetration depth of a low-energy projectile.

We conclude that the penetration depth of a low-energy impactor into a granular bed in microgravity seems to scale with the quantity $(\mu^{-1}(\rho_p/\rho_g)^{1/10}(D_p/d_g)^{2/15}H^{2/5})^{5/2}$ or equivalently $\mu^{-5/2}(\rho_p/\rho_g)^{1/4}(D_p/d_g)^{1/3}H$ (Figure~\ref{f:zmax_scaling}, right). This means that the larger the grain size in a regolith bed, the lower the penetration depth will be for the same size projectile. While the linear fits overlap very well, we observe a non-negligible remaining spread in the data. This could be due to a number of factors, including the width of the distribution in the grain size in each target.

%----------------------------------------------------------------- 
\section{Discussion}
\label{s:discussion}

\subsection{Mixed target material}

For nine of the 86 impacts performed, the target material was a mixture of the mm- and cm-sized grain populations. While we did not cover the entire parameter space for this type of target, these data points give an idea of how a mixed target behaves compared to monodisperse ones.

For the impact outcomes, we observed a high percentage of rolling on mixed targets, that is, 78\% compared to 57\% for cm-sized and 36\% for mm-sized targets (for a total of 51\% rolling observed as described in Section \ref{s:rolling2}). No ejecta blankets were observed, but this can be linked to the fact that the impact velocities into mixed targets never reached 1~m/s. Figure~\ref{f:masses} shows that the impact energies required for lifting grains from mixed targets ranged between the ones for mm- and cm-sized targets, supporting a simple scaling between the impact energy and the mass of lifted material as long as no ejecta blankets are created. 

Figure~\ref{f:COR} (and Section \ref{s:COR}) shows that the values and scatter in CORs are similar to monodisperse targets with a slightly higher average value and no noticeable dependence on the impact energy. The transfer from normal to tangential motion in mixed targets follows the same trend as monodisperse targets (Figure~\ref{f:COR_ratio}). We conclude that, with respect to rebound and rolling, mixed targets behave in the same way as monodisperse ones, with the only difference being that rolling is observed significantly more often.

\subsection{Wall effects}
\label{s:walls}

As described in Section \ref{s:HW} and listed in Table~\ref{t:runs}, we performed a number of impacts with much larger target samples: for 27 out of the 86 impacts, the whole bottom half of the experiment vacuum box was filled with material rather than contained in a 13$\times$13$\times$3 cm$^3$ tray. This was performed to exclude wall effects in the target response to our low-energy impacts. 

During data analysis, we systematically marked the large target samples to observe if the presence of a tray influenced the impact outcome. This is particularly relevant for cm-sized targets, as it has been reported that the presence of bottom and side walls in similar low-energy impacts can influence the experiment's outcome if the tray depth is shorter than a few grain diameters and the side walls are closer than five grain diameters to the impact site \citep{seguin2008}. Our analysis showed that, for the parameter space investigated here, no significant difference between tray and large samples could be seen, and data points were well mixed between the two populations. We therefore conclude that we can neglect the presence of tray side walls in our experiments.

\subsection{Cohesion in large regolith grains}

In Figure~\ref{f:outcomes}, we observe frequent bouncing of the projectile (triangles). Indeed, we find that 61\% of impacts result in a rebound, either normal to the target surface (R in Table~\ref{t:runs}), tangential to it (Ro), or a combination of both. Rebound without ejecta production (diamonds in Figure~\ref{f:outcomes}) was also very frequent (75~\% of the impacts not producing ejecta, 24~\% of the total number of impacts).

Just like with fine grain targets, the bouncing of the projectile is an indicator of a somewhat elastic behavior of the target. In \cite{brisset2018regolith}, it was only observed for impacts under microgravity conditions (compared to lunar-like 10$^{-2}g$ levels) indicating environments in which acceleration is too high successfully mask the inter-grain cohesion responsible for this bulk elasticity. While our experiment setup did not allow for an actual measure of the inter-grain cohesive forces, the high occurrence rate of bouncing events in our large grain data set shows that the target cohesion plays a non-negligible role in impact outcomes at the microgravity levels and impact speeds considered. We observe projectile bouncing even for the highest impact speeds investigated, showing that inter-grain cohesion between coarse grains remains significant even for several m/s impacts under the gravity conditions of small asteroids.

This indicates that surfaces composed of coarse grains on small bodies will display similar elastic behavior to ones composed of fine grains when impacted at low speeds. Any re-captured impact ejecta from a meteoritic impact or material lifted by surface activity on small asteroids will readily be accreted with very little to no new ejecta production.

\subsection{Comparing experimental data with numerical and spacecraft data}

Hayabusa 2's MASCOT landing on the surface of Ryugu in October 2018 \citep{scholten2019descent} presents a great opportunity to place our experimental findings in context with actual spacecraft data. While MASCOT is one order of magnitude larger than our projectiles and its impact onto the surface was oblique, a number of other parameters were similar to the ones explored in our experiment plan. In particular, the impact speed of 17~cm/s is in the low range investigated here and the size ratio between MASCOT and the boulders and coarse grains of Ryugu's surface ranges from 0.1 to 1, just like in our experiments. In addition, we can also compare our findings to numerical work performed specifically to simulate MASCOT's landing \citep{thuillet2018numerical,maurel2018numerical}.

\cite{scholten2019descent} describes the details of the descent and surface bouncing of MASCOT on Ryugu. The first contact point (CP1) is described as a multi-contact event resulting in MASCOT's bouncing off the surface again from a slightly different contact point (CP2). The impact at CP1 happened at 17.04~cm/s and the rebound speed from CP2 was 5.31~cm/s, for an overall COR of 0.31. In our data set, impact 2017-10-10-1 has very similar parameters with an impact speed at 17.7~cm/s and a COR of 0.34 into grains of sizes on the same order of the projectile. This impact resulted in rolling of the projectile on the target surface, which can be compared to a multi-contact event. If at CP2 MASCOT encountered the edge of the large boulder it was rolling on, it would have started a new free-fall phase with its rolling speed acquired during the impact at CP1, which is seen in the following free-fall trajectory to the third contact point (CP3). 

Coefficients of restitution at the next three contact points (CP3, 4, and 5) are increasingly high at 0.56, 0.63, and 0.98 with decreasing impact speeds (6.14, 3.80, and 2.33, respectively). While this follows the general trend, in our data of larger CORs at lower speeds we see that our predictions for these low speeds are lower, with \cite{brisset2018regolith} measuring a COR of about 0.3 at 4~cm/s into fine grains.

Compared to the MASCOT landing numerical simulations of \cite{thuillet2018numerical} and \cite{maurel2018numerical}, our experiments cover about the same size of target grains (1~cm), with material properties in the range of what they define as moderately frictional surfaces ($\mu$ = 0.53). One main difference lies in the size and shape of the projectile (2~cm sphere in our experiments, 30~cm cube in the MASCOT simulations), leading to a $D_p/d_g$ ratio about one order of magnitude larger. Nevertheless, we can compare our experimental results to normal impacts (0$\degree$ impact angle) in the simulations.

The collision outcome parameter easiest to compare to our experiments is the COR, which is named "linear COR" in these numerical works. The first thing we notice is that the presented simulations are sensitive to the regolith bed depth when this one is shallower than 30~cm. Our experiments show a very different behavior, in which target samples of 3 and 30~cm depths do not display noticeable differences (see Section \ref{s:walls}). In addition, strong inter-grain friction (named "gravel" in their work) seems to match our experimental results better than the moderately frictional material, despite the fact that the material properties we measure for our target material match a moderate friction.

At about 20~cm/s, we record CORs ranging between 0.1 and 0.35. In \cite{thuillet2018numerical}, normal impacts into a 15~cm highly frictional regolith bed result in CORs between 0.3 and 0.35, values that are well reproduced by our experiments. Moderately frictional material for the same impact conditions results in lower CORs between 0.01 and 0.1, the higher end of which still matches the lower end of our measurements. In \cite{maurel2018numerical}, normal impacts into moderately frictional material results in projectile embedding. In similar impact conditions, we did see a few instances of projectile embedding (see Figure~\ref{f:outcomes}, circles close to 20~cm/s impact speed), but the significantly predominant outcome of impacts at similar speeds was rebound with CORs concentrating between 0.2 and 0.3. This better matches the \citep{maurel2018numerical} numerical results obtained for a 15~cm highly frictional regolith bed, where the computed COR is around 0.3.

Another parameter we can compare our experimental results to is the maximum penetration depth calculated for the MASCOT impact simulations. If we apply MASCOT parameters to the scaling law we derived in Section \ref{s:penetration_depth}, we can calculate an empirically derived maximum penetration depth for the payload. For this purpose, we chose to use the moderate friction coefficient defined in the simulations $\mu$ = tan(28$\degree$) =  0.53. Furthermore, MASCOT's density is calculated from its volume (29.0 $\times$ 27.5 $\times$ 19.5 cm$^3$) and mass (10~kg) as $\rho_p$ = 0.643 g/cm$^3$. The bulk density of the target in the simulations is set to $\rho_g$ = 1.28 g/cm$^3$. For a flat impact, the projectile diameter can be taken as MASCOT's largest cross-section diagonal, $D_p$ = 40~cm. The peak of the target grain size distribution in the simulations is set to $d_g$ = 1~cm. The impact speed is considered to be 19~cm/s for a 1$g$ equivalent drop height of H = 1.84~mm. These parameters result in a maximum penetration depth of 7.36~cm. We note that this is about a factor of 2 shallower than the depths calculated for moderate friction regolith in \cite{thuillet2018numerical} and \cite{maurel2018numerical} (about 15~cm in both). If we use the higher friction coefficient ($\mu$ = tan(38.5$\degree$) =  0.80), which better matches our experimental COR values, we obtain a penetration depth of 2.63~cm. This is still a factor of 3 to 4 shallower than the results of \cite{thuillet2018numerical} (about 10~cm), but closer to results obtained by \cite{maurel2018numerical} (about 3.51~cm).

For a possible reconciliation between simulation and experimental results, we note that the frictional material properties of our target sample were measured in the laboratory under 1$g$ conditions. Given the target behavior sensitivity to gravity levels \citep{brisset2018regolith}, it is very possible that the inter-grain friction is different in microgravity. In particular, if gravity is masking inter-grain cohesion forces, angle of reposes are expected to be lower in 1$g$ compared to $< 10^{-4}g$. This could explain why our COR measurements better match the highly frictional materials in the simulations even though we measured moderate friction for the target material in the laboratory.

Another factor in the COR differences between simulations and experiments could be the projectile-to-target size ratio, which is a factor of 10 higher in the simulations compared to the experiments. In \cite{thuillet2019numerical}, the same team studied the armoring effect of the presence of a boulder (of which the dimensions are roughly similar to those of MASCOT ($D_p/d_g$ = 1)) at the payload's impact site. Interestingly, CORs for a boulder flush with the regolith surface have values around 0.35, which are similar to the experimental values we observed in our work. Even though we do not see a grain size dependence in our COR measurements (see Section \ref{s:COR2}), this could indicate a possible reconciliation between our data and the numerical outcomes.

%-----------------------------------------------------------------
\section{Conclusion}
\label{s:conclusion}

We performed low-velocity impacts into large grain targets (mm- and cm-sized grains) under microgravity conditions. The laboratory PRIME drop-tower setup allowed us to record impacts of cm-sized projectiles into large grain targets prepared from a UCF-1 asteroid simulant in acceleration environments $< 10^{-4}g$. The main results of our data analysis can be summarized as follows:

\begin{itemize}
\item \textit{Impact Outcomes\emph{:}} similarly to microgravity impacts into fine grains \citep{brisset2018regolith}, we observed a variety of impact outcomes, noticeably including a projectile rebound off the target's surface after impact (61 \% of the observed impacts). We also observed a new outcome that was not seen for fine grain targets: the rolling of the projectile at the surface of the target after initial impact (see Figure~\ref{f:rolling}). This outcome was very frequent, occurring for about half the impacts (overlapping with rebound on occasion). We note that the frequent observation of projectile rolling is suspected to be due to the projectile and regolith grains being on the same order in size.

\item \textit{Ejecta Speeds\emph{:}} the measurement of ejecta speeds showed the limitations of our experimental setup for impact speeds below about 40~cm/s. For higher impact speeds, we were able to identify a clear trend in average ejecta speeds both with the impact speed and the target grain size (Figure~\ref{f:espeed_ispeed}). We were able to scale the ejecta speeds to the quantity $(D_p/d_g)^{1/3}v_i^{1/2}$ (Figure~\ref{f:espeed_scaling}).

\item \textit{Ejecta Masses\emph{:}} estimations for the ejected mass upon impact showed a simple scaling between the impact energy and the mass of the lifted material as long as no ejecta curtain is created. This trend is reversed for the lifting of ejecta blankets, indicating that the target material cohesion becomes apparent for larger impact energies and that larger grains present a looser surface than smaller ones.

\item \textit{Coefficient of Restitution\emph{:}} for impacts leading to rebound and rolling of the projectile, we measured the associated COR. Our data set showed a relatively low dependence of the COR on the target grain size (average values around 0.16, similar to results in fine grain targets. We see a dependence on the impact energy with an exponential fit index of -0.2 for normally rebounding projectiles (Figure~\ref{f:COR_norolling}).

\item \textit{Projectile Penetration Depth\emph{:}} For a number of impacts, we also measured the maximum penetration depth of the projectile into the target material ($z_{max}$). Combining the data sets for fine grains \citep{brisset2018regolith}, mm-, and cm-sized grains, we were able to scale $z_{max}$ with the quantity  
$\mu^{-5/2}(\rho_p/\rho_g)^{1/4}(D_p/d_g)^{1/3}H$ (see \ref{s:penetration_depth}), indicating that the larger the grain size in a regolith bed, the lower the penetration depth will be for the same size projectile.

\end{itemize}

Combining our data set with the one analyzed in \cite{brisset2018regolith}, we learn that grain size in the regolith at the surface of a small body (e.g., an asteroid) impacts the ejecta mass and average speed, as well as the capacity of a low-velocity projectile to enter the surface. The rebound coefficient of restitution however, is insensitive to grain size. These findings could be of particular interest for ongoing or upcoming exploration missions to small asteroids, for which interactions with a surface covered in coarse grains has been observed or is planned.

\begin{acknowledgements}
      This work was funded by the Florida Space Institute with undergraduate student support partially funded by the EXCEL mentoring program of the University of Central Florida. We would like to thank Josh Colwell for providing access to his laboratory drop tower.
\end{acknowledgements}

% WARNING
%-------------------------------------------------------------------
% Please note that we have included the references to the file aa.dem in
% order to compile it, but we ask you to:
%
% - use BibTeX with the regular commands:
%   \bibliographystyle{aa} % style aa.bst
%   \bibliography{Yourfile} % your references Yourfile.bib
%
% - join the .bib files when you upload your source files
%-------------------------------------------------------------------

\bibliographystyle{aa}
\bibliography{Literatur.bib}

%
%-------------------------------------------------------------
%          For the appendices, table longer than a single page
%-------------------------------------------------------------

% Table will be print automatically at the end of the document, 
% after the whole appendices

\begin{appendix} %First appendix
\section{Experiment Run Table}

% In the appendices do not forget to put the counter of the table 
% as an option

\longtab[1]{
\begin{longtable}{|c|c|c|c|c|l|c|c|}

\caption{List of low-velocity impacts into coarse regolith performed under microgravity conditions for the present investigation. The tray column indicates the use of a target tray (1) or not (0). The target material listed are mixed (0), mm-sized (1), or cm-sized (2) grains. The impact outcomes are sorted as Embedding (E), Rebound (R), Rolling (Ro), or a combination of them. For rolling or rebound outcomes, the projectile coefficient of restitution (COR) is given. An estimation of the ejected mass is also listed: (0) no ejecta; (1) the ejected mass is much lower than the projectile mass; (2) the ejected mass is on the order of the projectile mass; (3) the ejected mass is much higher than the projectile mass.}
\label{t:runs} \\
\hline Drop ID & Tray & Target & Proj. Mass [g] & Impact Speed [cm/s] & Outcome & COR & Ejected Mass \\ \hline 
\endfirsthead

\multicolumn{8}{c}%
{{\bfseries \tablename\ \thetable{} -- continued from previous page}} \\
\hline Drop ID & Tray & Target & Proj. Mass [g] & Impact Speed [cm/s] & Outcome & COR & Ejected Mass \\ \hline 
\endhead

\hline \multicolumn{8}{|r|}{{Continued on next page}} \\ \hline
\endfoot

\hline \hline
\endlastfoot

2017-08-24-1 & 1 & 0 & 4.8 & 34.2 & Ro & 0.0469 & 1 \\ \hline
2017-08-29-1 & 1 & 0 & 9.81 & 80.6 & Ro & 0.1483 & 1 \\ \hline
2017-08-30-1 & 1 & 0 & 9.81 & 59.5 & Ro & 0.2023 & 1 \\ \hline
2017-08-30-2 & 1 & 0 & 9.81 & 60.6 & Ro & 0.3479 & 2 \\ \hline
2017-08-31-1 & 1 & 0 & 4.8 & 33.9 & Ro & 0.1963 & 0 \\ \hline
2017-09-01-1 & 1 & 1 & 4.8 & 33.5 & Ro & 0.2294 & 0 \\ \hline
2017-09-05-1 & 1 & 1 & 4.8 & 37.4 & Ro & 0.2228 & 1 \\ \hline
2017-09-07-1 & 1 & 1 & 4.8 & 32.6 & R  & 0.1850 & 0 \\ \hline
2017-09-18-1 & 1 & 1 & 9.81 & 66.9 & E & 0 & 2 \\ \hline
2017-09-18-2 & 1 & 1 & 9.81 & 71.2 & Ro & 0.0642 & 1 \\ \hline
2017-09-19-1 & 1 & 1 & 9.81 & 36.6 & R  & 0.1963 & 1 \\ \hline
2017-09-21-1 & 1 & 1 & 30.75 & 14.4 & E & 0 & 2 \\ \hline
2017-09-21-2 & 1 & 1 & 30.75 & 20.9 & E & 0 & 2 \\ \hline
2017-09-25-1 & 1 & 1 & 30.75 & 19.2 & ERo & 0 & 2 \\ \hline
2017-09-26-1 & 1 & 1 & 30.75 & 9.7 & ERo & 0 & 1 \\ \hline
2017-09-26-2 & 1 & 1 & 30.75 & 15.2 & Ro & 0.1023 & 0 \\ \hline
2017-09-27-1 & 1 & 1 & 30.75 & 13.6 & Ro & 0.0639 & 0 \\ \hline
2017-09-28-1 & 1 & 0 & 30.75 & 8.5 & ERo & 0 & 0 \\ \hline
2017-09-29-1 & 1 & 0 & 30.75 & 11.1 & E & 0 & 0 \\ \hline
2017-10-02-1 & 1 & 0 & 30.75 & 13.4 & Ro & 0.2176 & 0 \\ \hline
2017-10-05-1 & 1 & 0 & 30.75 & 9.1 & R  & 0.1427 & 0 \\ \hline
2017-10-06-2 & 1 & 2 & 30.75 & 9 & E & 0 & 0 \\ \hline
2017-10-06-1 & 1 & 2 & 30.75 & 8.8 & Ro & 0.2248 & 0 \\ \hline
2017-10-09-1 & 1 & 2 & 30.75 & 22.6 & E & 0 & 0 \\ \hline
2017-10-09-2 & 1 & 2 & 30.75 & 16.2 & E & 0 & 0 \\ \hline
2017-10-09-3 & 1 & 2 & 30.75 & 7.8 & Ro & 0.0122 & 0 \\ \hline
2017-10-10-1 & 1 & 2 & 30.75 & 17.7 & Ro & 0.3410 & 0 \\ \hline
2017-10-11-1 & 1 & 2 & 4.8 & 24.5 & Ro & 0.2141 & 0 \\ \hline
2017-10-11-2 & 1 & 2 & 30.75 & 57.3 & Ro & 0.2059 & 1 \\ \hline
2017-10-12-1 & 1 & 2 & 4.8 & 31.6 & Ro & 0.2420 & 0 \\ \hline
2017-10-12-2 & 1 & 2 & 4.8 & 21.1 & Ro & 0.0736 & 0 \\ \hline
2017-10-13-1 & 1 & 2 & 9.81 & 50.1 & R  & 0.2646 & 0 \\ \hline
2017-10-16-1 & 1 & 2 & 9.81 & 31.9 & Ro & 0.1015 & 0 \\ \hline
2017-10-16-2 & 1 & 2 & 9.81 & 48 & Ro & 0.0828 & 0 \\ \hline
2018-01-09-1 & 1 & 2 & 30.75 & 179.1 & R  & 0.1090 & 3 \\ \hline
2018-01-10-1 & 1 & 2 & 30.75 & 19.4 & R  & 0.1128 & 1 \\ \hline
2018-01-10-2 & 1 & 2 & 30.75 & 131.8 & R  & 0.1897 & 3 \\ \hline
2018-01-10-3 & 1 & 2 & 30.75 & 120.7 & Ro & 0.0566 & 3 \\ \hline
2018-01-11-1 & 1 & 2 & 30.75 & 89.7 & RRo & 0.1071 & 1 \\ \hline
2018-01-16-1 & 1 & 2 & 30.75 & 64.9 & R  & 0.0308 & 3 \\ \hline
2018-01-17-2 & 1 & 2 & 30.75 & 80.4 & E & 0 & 0 \\ \hline
2018-01-17-1 & 1 & 2 & 30.75 & 75 & Ro & 0.1929 & 3 \\ \hline
2018-01-19-1 & 1 & 2 & 30.75 & 27.3 & Ro & 0.3763 & 0 \\ \hline
2018-01-22-1 & 1 & 2 & 30.75 & 23.3 & E & 0 & 0 \\ \hline
2018-01-22-2 & 1 & 2 & 30.75 & 24.8 & E & 0 & 0 \\ \hline
2018-01-23-2 & 1 & 2 & 30.75 & 157.2 & R  & 0.0546 & 3 \\ \hline
2018-01-23-1 & 1 & 2 & 30.75 & 164.6 & Ro & 0.1394 & 0 \\ \hline
2018-01-24-1 & 1 & 2 & 30.75 & 139 & R  & 0.1860 & 3 \\ \hline
2018-01-24-2 & 1 & 2 & 30.75 & 36 & RRo & 0.4281 & 3 \\ \hline
2018-01-29-1 & 1 & 2 & 30.75 & 120 & Ro & 0.1665 & 2 \\ \hline
2018-01-31-1 & 1 & 2 & 30.75 & 115.6 & Ro & 0.1313 & 2 \\ \hline
2018-02-01-1 & 1 & 2 & 30.75 & 95.2 & Ro & 0.1402 & 2 \\ \hline
2018-02-06-1 & 1 & 1 & 30.75 & 112.1 & E & 0 & 3 \\ \hline
2018-02-07-1 & 1 & 1 & 30.75 & 126.9 & R  & 0.0327 & 3 \\ \hline
2018-02-08-1 & 1 & 1 & 30.75 & 129.1 & E & 0 & 3 \\ \hline
2018-02-13-1 & 1 & 1 & 30.75 & 162.6 & RRo & 0.1063 & 3 \\ \hline
2018-02-14-1 & 1 & 1 & 30.75 & 162.4 & RRo & 0.0450 & 3 \\ \hline
2018-02-16-1 & 1 & 1 & 30.75 & 165.1 & RRo & 0.0550 & 3 \\ \hline
2018-02-20-1 & 1 & 1 & 30.75 & 166.9 & E & 0 & 3 \\ \hline
2018-05-14-1 & 0 & 2 & 30.75 & 23.8 & ERo & 0 & 1 \\ \hline
2018-05-29-1 & 0 & 2 & 30.75 & 95.1 & E & 0 & 2 \\ \hline
2018-06-04-1 & 0 & 2 & 30.75 & 112 & Ro & 0.1191 & 2 \\ \hline
2018-06-05-1 & 0 & 2 & 30.75 & 126.7 & Ro & 0.1247 & 2 \\ \hline
2018-06-06-1 & 0 & 2 & 30.75 & 165.7 & Ro & 0.0696 & 2 \\ \hline
2018-06-08-1 & 0 & 2 & 30.75 & 167.9 & Ro & 0.1026 & 1 \\ \hline
2018-06-12-1 & 0 & 2 & 30.75 & 57 & R  & 0.1267 & 2 \\ \hline
2018-06-13-1 & 0 & 1 & 30.75 & 68.2 & E & 0 & 3 \\ \hline
2018-06-15-1 & 0 & 1 & 30.75 & 31.1 & E & 0 & 2 \\ \hline
2018-06-19-1 & 0 & 1 & 30.75 & 110.8 & E & 0 & 3 \\ \hline
2018-06-21-1 & 0 & 1 & 30.75 & 178.1 & E & 0 & 3 \\ \hline
2018-06-25-1 & 0 & 1 & 30.75 & 184.2 & E & 0 & 3 \\ \hline
2018-06-27-1 & 0 & 1 & 30.75 & 165.1 & E & 0 & 3 \\ \hline
2018-06-27-2 & 0 & 1 & 30.75 & 142.4 & E & 0 & 3 \\ \hline
2018-06-28-1 & 0 & 1 & 30.75 & 95.3 & E & 0 & 3 \\ \hline
2018-07-02-1 & 0 & 1 & 30.75 & 116.4 & E & 0 & 3 \\ \hline
2018-08-03-1 & 0 & 1 & 30.75 & 28.3 & E & 0 & 2 \\ \hline
2018-08-10-1 & 0 & 1 & 30.75 & 27.9 & E & 0 & 3 \\ \hline
2018-08-22-1 & 0 & 2 & 30.75 & 19.5 & R  & 0.1770 & 0 \\ \hline
2018-08-27-1 & 0 & 2 & 30.75 & 28.3 & Ro & 0.1745 & 0 \\ \hline
2019-03-06-1 & 0 & 2 & 30.75 & 59.3 & E & 0 & 2 \\ \hline
2019-03-08-1 & 0 & 2 & 30.75 & 45.3 & R  & 0.1737 & 2 \\ \hline
2019-03-25-1 & 0 & 2 & 30.75 & 110.2 & E & 0 & 1 \\ \hline
2019-04-08-1 & 0 & 2 & 30.75 & 171.1 & RRo & 0.0734 & 2 \\ \hline
2019-04-12-1 & 0 & 1 & 30.75 & 72 & Ro & 0.0656 & 3 \\ \hline
2019-04-15-1 & 0 & 1 & 30.75 & 71.5 & E & 0 & 2 \\ \hline
2019-04-17-1 & 0 & 1 & 30.75 & 62.1 & E & 0 & 3 \\ \hline
\end{longtable}
%\end{center}
}% End longtab

\end{appendix}

\end{document}